\documentclass[a4paper,12pt]{article}

\usepackage{amsmath,amssymb,amsfonts,epsfig,graphicx,colordvi}

\usepackage{colordvi}

\title{The magnetization plateaus of the ferro and anti-ferro 
spin-1 classical models with $S_z^2$ term}

\author{
 S.M de Souza$^{1}$ and  M.T. Thomaz$^{2}\footnote{Corresponding author: mtt@if.uff.br}$
\vspace{0.25cm} \\
\small\it $^{1}$ Departamento de Ci\^encias Exatas, Universidade Federal de Lavras,\\ 
\small\it Caixa Postal 3037, CEP 37200-000, Lavras-MG,  Brazil.
\vspace{0.25cm}\\
\small\it $^{2}$Instituto de F\'{\i}sica, Universidade Federal Fluminense,\\ 
\small\it Av. Gal. Milton Tavares de Souza s/n$^{\textit o}$, 
CEP 24210-346, Niter\'oi-RJ, Brazil. 
}

\begin{document}

\maketitle

\begin{abstract}

We study in detail   the exact thermodynamics of the one-dimensional
standard and staggered spin-1 Ising models  with a 
single-ion anisotropy term in the presence of a longitudinal
magnetic field at low temperatures. The results are valid for the 
ferromagnetic and anti-ferromagnetic (AF) models and for positive
and negative values of the crystal field for $T>0$. Although the
excited states of the ferro and anti-ferro models are highly
degenerate, we show that the temperature 
required for reaching the first excited state 
in  the classical spin-1 ferro model
gives a scale of temperature that permits 
fitting the $z$-component of the magnetization only by the
contribution of two ground states of the model. This
approximation is not true for the equivalent AF
function due to the fact that the AF model is gapless
along the lines separating the phases in its  phase
diagram at $T=0$. We relate the number of plateaus in the
magnetization of each  model to their respective  phase 
diagrams at $T=0$.  The specific heat per site of the AF 
model distinguishes, at low temperature,  the transitions 
$\mbox{A} \rightleftharpoons \mbox{E}$ and 
$\mbox{G} \rightleftharpoons \mbox{E}$ 
as the external magnetic field is varied. The exact Helmholtz  free
energy of the classical spin-1 model is mapped onto the 
equivalent function  of the ionic limit of the $1D$ 
extended Hubbard model by proper transformations.

\end{abstract}

\vfill


\newpage


\section{Introduction}  \label{sec_1}

Exactly solvable  models help us having insights on
more complex systems. In one space dimension ($1D$), the
thermodynamics of the spin-$\frac{1}{2}$ Ising model in the presence
of a longitudinal magnetic field has been solved  in the
whole interval of temperature using the density matrix
approach\cite{kramers1,kramers2,baxter}. To do so, it is essential the
commutative nature of all operators in its Hamiltonian. 
This model is usually called a classical system.

Cold atoms have made possible simulations of spin models. 
Recently Simon {\it et al}.\cite{simon}
simulated a one-dimensional spin-$1/2$ model in the 
presence of a magnetic field with longitudinal and
transverse components by using  a Mott insulator of
spinless bosons in a tilted optical lattice. The device
was used to study the phases of this  classical model at 
low temperatures. These optical devices have permitted 
the experimental study of properties of chain models,
including the spin models.

The spin-1 Ising model with single-ion anisotropy 
term, the Blume\cite{blume}-Capel\cite{capel} model,
in the presence of a longitudinal magnetic field, has
also a classical nature. This fact permits us to apply 
the transfer matrix method\cite{kramers1,kramers2,baxter}  to solve it.
With this approach, Aydiner and Aky\"uz \cite{aydiner} 
obtained the numerical solution of the $1D$ spin-1 
anti-ferromagnetic (AF) Ising model with a single-ion 
anisotropy term in the presence of a longitudinal 
magnetic field. They studied the magnetic and thermal 
behavior of the model at very low temperatures. 
In 2003 Chen {\it et al}.\cite{chen} applied the classical 
Monte Carlo (CMC)  technique to the numerical calculation of 
the  phase diagram at $T=0$ of the spin-$1$ 
anti-ferromagnetic Ising model in the presence of a 
single-ion anisotropy term for positive crystal field.
The CMC was applied also to the study of  magnetization 
plateaus of this model at low temperature.

In 2005 Mancini\cite{mancini} used the mapping between 
the $1D$ extended Hubbard model 
in the ionic limit  (i.e., $|t| \ll |U|, |V|$, in which
$t$ is the hopping exchange,  $U$ 
is the on-site Coulomb interactions and $V$ is the inter-site 
Coulomb interaction) and the spin-1 Ising 
model with a single-ion anisotropy term in the presence of a 
longitudinal magnetic field to write the exact Helmholtz 
free energy (HFE) of the latter\cite{mancini} as a set 
of coupled equations. The hierarchy of the equations of 
motion is closed, and they  can be solved numerically.
In Ref.\cite{mancini2008} they have extended the approach to 
include a biquadratic nearest-neighbour interaction term,
numerically solving  the coupled 
equations and presenting the behavior of some
thermodynamic functions at low temperatures.
We derived in Ref.\cite{winder} the exact expression of 
the HFE of the $1D$ spin-1 Ising model with the
$(S_z)^2$ term  in the absence of a magnetic field.
This HFE has a simple analytic expression  valid for
the ferromagnetic and AF models in the
whole range  of temperature. The absence of a magnetic 
field in this solution 
prevents accessing some information about 
the system, e.g.: information about the magnetization  
and  the susceptibility of the model.

All the previously mentioned papers on the one-dimensional 
spin-1 Ising model seem to be unaware  of the nice work 
by Krinsky and Furman\cite{krinsky} in 1976, which
presents the calculation of
the exact HFE for the $1D$ ferromagnetic 
spin-1 Ising model with a single-ion anisotropy term, a 
biquadratic nearest-neighbour interaction term  and a 
nonsymmetric term, in the presence of a longitudinal  
magnetic field. 
More recently Litaiff {\it et al}. also applied the 
transfer-matrix technique to calculate the exact expression
of the HFE of this model. The authors did not use this 
thermodynamic function to explore its low 
temperature behavior.

Although the $1D$ spin-$S$ Ising model  with the 
single-ion anisotropy  term in the presence of a magnetic
field is called a classical model, it has a quantum nature
that manifests itself as plateaus in the 
$z$-component of the  magnetization at very low temperature.
The existence of those plateaus has been demonstrated in 
a number of $1D$ models\cite{ekiz}  and experimentally
measured in some materials\cite{narumiPhysB,narumiJMMM,goto}
described by $1D$ spin-models. 
Magnetization plateaus in the the AF  spin-1 Ising model with 
positive crystal field in the presence of a longitudinal
magnetic field have been obtained numerically  in 
Refs.\cite{aydiner,chen,mancini2008}. The authors of 
Ref.\cite{krinsky} mentioned the discontinuous change
of the magnetization of the ferromagnetic model, but their
focus was on the exact renormalization group  of the model
and  the discussion of its critical points.

\vspace{0.3cm}

The study of the simple one-dimensional spin-1 Ising model
helps to understand the origin of plateaus in 
the magnetization function.  
In the present communication we present the exact 
analytic  expressions of the HFE's of the standard and
the staggered versions of the $1D$ spin-1 Ising model 
with the $(S_i^z)^2$ term in the presence 
of a longitudinal magnetic field, valid 
for $T>0$ in section \ref{sec_2}. Our solutions apply equally well to the 
ferromagnetic and to the AF models of both versions,
extending the validity of the solution derived
in Ref.\cite{krinsky} for the ferromagnetic model.
We study the $z$-component  of the magnetization  
and the entropy per site at very low temperatures
for the ferro and AF versions of the classical spin-1 model
in sections \ref{sec_3} and \ref{sec_4}, respectively. 
We relate the number of plateaus in the magnetization  to
the number of phases in the ground state of the
model at $T=0$.
Finally in section \ref{sec_5} we present our
conclusions. More detailed calculations are shown in 
Appendices \ref{Apend_A} and \ref{Apend_B}.


\section{The Hamiltonians of the classical spin-1 models
in the presence of a longitudinal magnetic field}  \label{sec_2}

The Hamiltonian of the one-dimensional of the classical spin-1 Ising model
with a single-ion anisotropy term, the Blume-Capel model, in the
presence of a longitudinal magnetic field is\cite{blume,capel}

\begin{eqnarray} \label{1}
{\mathbf H}^{\prime} &=& \overset{N}{\underset{i=1} \sum} \;\;
\left[ J S_i^z S_{i+1}^z  - h^{\prime} S_i^z 
+ D^{\prime} (S_i^z)^2 \right]   \nonumber \\
&=& \overset{N}{\underset{i=1} \sum} \;\; {\bf H}_{i, i+1} ,
\end{eqnarray} 

\noindent  where $S_i^z$ is the $z$- component  of the
spin-1  operator  with norm: $||\vec{S}||^2 =  2$. 
$J$ is the exchange strength and it can 
have either a negative value (ferromagnetic model) 
or a positive value (AF model). The crystal field $D$ 
can assume positive, negative or null values. The model
satisfies spacial periodic boundary condition in a chain 
with $N$ sites. The external magnetic field 
is oriented along the easy-axis $z$. In 
this paper we use natural units, $e= m= \hbar =1$. 
Comparing our Hamiltonian (\ref{1}) to the Hamiltonian 
(25) in Ref.\cite{litaiff} we have
$J = J^{*}$, $h^{\prime} = \mu_B H^{*}$ and 
$D^{\prime} = D^{*}$, in which  $J^{*}$,
$H^{*}$ and $D^{*}$  
 are the parameters of the Hamiltonian 
of the single spin-1 Ising model in Ref.\cite{litaiff}.

Mancini\cite{mancini} showed that by the
substitution $S_i^z = {\mathbf n}_i - {\mathbf 1}_i$ (where 
${\mathbf 1}_i$ is the identity 
operator  at the $i$-th site, in which
${\mathbf n}_i = {\mathbf n}_{i, \uparrow} + {\mathbf n}_{i, \downarrow}$,\  \ 
${\mathbf n}_{i, \sigma} \equiv {\mathbf c}_{i, \sigma}^{\dag} {\mathbf c}_{i, \sigma} $ with 
$ \sigma \in \{ \uparrow, \downarrow\}$, 
${\mathbf c}_{i, \sigma}^{\dag}$
is the fermionic creation
operator of an electron at the $i$-th site with
the spin component $\sigma$ and
${\mathbf c}_{i, \sigma}$ is the corresponding
destruction operator,
the Hamiltonian (\ref{1}) is mapped onto the ionic limit of
the Hamiltonian of the  1D extended Hubbard model
in the presence of a chemical potential, that is 

\begin{eqnarray}  \label{2}
{\mathbf H}^{\prime} = {\mathbf H}_{Hub} 
      + N (J + h^{\prime} + D^{\prime}) \; {\mathbf 1} ,
\end{eqnarray}

\noindent where\cite{luz,moutinho}

\begin{eqnarray} \label{3}
{\mathbf H}_{Hub} = \sum_{i=1}^{N} \; \;  \left(
U {\mathbf n}_{i, \uparrow} {\mathbf n}_{i, \downarrow} 
+ 2V {\mathbf n}_i \, {\mathbf n}_{i+1} 
- \mu {\mathbf n}_i  \right) .
\end{eqnarray}

\noindent  The last equality is valid for $U = 2 D^{\prime}$,
$V = \frac{J}{2}$ and $\mu = 2J + h^{\prime} + D^{\prime}$.
As a consequence of relation (\ref{2}) we have, for these 
values of parameters, that ${\mathcal W}_{Hub} (U, V, \mu; \beta)  
= {\mathcal W}_1 (J, h^{\prime}, D^{\prime}; \beta)
- (J + h^{\prime} + D^{\prime})$, where ${\mathcal W}_{Hub}$
(${\mathcal W}_1$) is the HFE of the extended Hubbard model
in the ionic limit (the  HFE of the spin-1 Ising model 
(\ref{1})). One is reminded that $\beta = \frac{1}{kT}$, 
where $k$ is the Boltzmann's constant and $T$ is the absolute 
temperature in kelvin.

\vspace{0.3cm}
 
The Hamiltonian (\ref{1}) in the presence of a staggered
longitudinal magnetic field is

\begin{eqnarray} \label{4}
{\mathbf H}_{stag}^{\prime} &=& \overset{2M}{\underset{i=1} \sum} \;\;
\left[ J S_i^z S_{i+1}^z  - h^{\prime} \; (-1)^i \;S_i^z 
+ D^{\prime} (S_i^z)^2 \right]  .
\end{eqnarray} 

\noindent The Hamiltonian continues to satisfy space 
periodic condition, but now we have an even number of 
sites in the chain: $N= 2M$. 

Applying the method presented in Ref.\cite{chain_m} 
we calculated in Ref.\cite{isingSz2} the $\beta$-expansion of the
HFE of the one-dimensional spin-$S$ Ising model, with single-ion 
anisotropy term, in the presence of a longitudinal magnetic
field,  ${\mathcal W}_S (J, h^{\prime}, D^{\prime}; \beta)$,
up to order $\beta^{17}$, in the thermodynamic limit. 
The $\beta$-expansion of the 
HFE of the staggered Hamiltonian (\ref{4}), with $D^{\prime} =0 $, 
${\mathcal W}_S^{stag} (J, h^{\prime}, 0; \beta)$,
was derived in Ref.\cite{physA2011} up to the same order in 
$\beta$  also for $M\rightarrow \infty$. 
Unfortunately those expansions do not permit us to study 
the thermodynamics of the standard and the staggered $1D$
 spin-1 Ising model close to $T=0$. In our web page\footnote
{ Our web page: {\tt http://www.proac.uff.br/mtt}.}
we provide the data files with  the quantum (arbitrary spin-$S$) and
the classical HFE's of the normalized Hamiltonians (\ref{1}) and 
(\ref{4})  up to order $\beta^{17}$ for both versions 
(standard and staggered) of the model.

In the present work we apply the transfer matrix
method\cite{baxter} together with the $\beta$-expansion of the 
function ${\mathcal W}_1 (J, h^{\prime}, D^{\prime}; \beta)$ 
obtained from the results of Ref.\cite{isingSz2} with $S=1$
to calculate the exact  HFE of  the Hamiltonian (\ref{1})
in the thermodynamic limit ($N \rightarrow \infty$),
valid for $T \stackrel{>}{_{\sim}} 0$.  The expression
of ${\mathcal W}_1^{stag} (J, h^{\prime}, D^{\prime}; \beta)$
is obtained by using a  well known result 
in the literature\cite{physA2011},  namely,
${\mathcal W}_S^{stag} (J, h^{\prime}, D^{\prime}; \beta) =
{\mathcal W}_S ( - J, h^{\prime}, D^{\prime}; \beta)$, with
$S= \frac{1}{2}, 1 , \frac{3}{2}, \cdots$.  This equality   is 
valid for $T\in [0, \infty)$.

\vspace{0.5cm}

In order to calculate the exact function 
${\mathcal W}_1 (J, h^{\prime}, D^{\prime}; \beta)$, valid for
$T>0$   and in the limit of $N \rightarrow \infty$,
using the transfer matrix method, we rewrite Hamiltonian
(\ref{1}) as a symmetric operator in the $i$-th and $(i+1)$-th 
sites, that is, 

\begin{eqnarray} \label{5}
{\bf H}_{i, i+1}^{(S)} =  J S_i^z S_{i+1}^z  - h S_i^z - h S_{i+1}^z
+ D (S_i^z)^2 + D (S_{i+1}^z)^2. 
\end{eqnarray} 

By comparing eqs.(\ref{1}) and (\ref{5}), we verify 
that the  $\beta$-expansion of the HFE of this model,  
in terms of the parameters $J$, $h$ and $D$ in Hamiltonian 
(\ref{5}) may be obtained by applying the following change of variables: 
$h^{\prime} \rightarrow 2h$ and $D^{\prime} \rightarrow 2D$, 
in the expansion of Ref.\cite{isingSz2} with $S=1$.

\vspace{0.5cm}

In appendix \ref{Apend_A} we calculate the three  roots 
of the third degree equation derived from the transfer matrix 
method for the classical spin-1 model with a single-ion anisotropy 
term in the presence of a longitudinal magnetic field. The root with
the largest modulus gives the expression of the HFE of the 
model (\ref{1})/(\ref{5}) (see eq. (\ref{A.4})).

In order to verify which root, eqs.(\ref{A.8a}) - (\ref{A.8c}), corresponds
to the eigenvalue $\lambda_1$ of matrix {\bf U}, 
assumed to be the root with the largest modulus,
we calculate the $\beta$-expansion of the functions 
$- \frac{1}{\beta} \ln[s_i]$, $i = 1, 2, 3$, and compare each one
with the expansion of the HFE of Ref.\cite{isingSz2} with $S=1$ 
(having made the change of variables $h^{\prime} = 2 h$
and $D^{\prime} = 2 D$ in the $\beta$-expansion of 
Ref.\cite{isingSz2}).  By direct comparison of the expansions, 
we obtain for  finite value of $\beta$, up to order $\beta^{10}$, that
the root $s_1$, eq.(\ref{A.8a}), is the eigenvalue $\lambda_1$, that
has the largest modulus among the eigenvalues of matrix {\bf U}. Our 
analysis  is valid for finite value of $\beta$, which excludes 
the value $T=0$.

Our solution $s_1$ is equal to the eigenvalue of the cubic equation 
with  the largest modulus calculated in Ref.\cite{krinsky} for the
ferromagnetic model ($J<0$). The way we write the 
solutions (\ref{A.9a}) - (\ref{A.9c}) avoids the necessity 
of defining a cut in the complex plane for their calculation. Our results 
extended those derived in Ref.\cite{krinsky} to include the
HFE of the AF spin-1 Ising model in the presence of
a longitudinal magnetic field.

The exact HFE of the standard one-dimensional spin-1 Ising 
model with the  single-ion anisotropy term in the presence 
of a longitudinal  magnetic field $h$ is

\begin{eqnarray} \label{6}
{\mathcal W}_1 (J, h, D; \beta) =  - \frac{1}{\beta}
\;\; \ln\left[2  \sqrt{- \tilde{Q}} \;\; \cos(\frac{\theta}{3}) 
    + \frac{P}{3} \right]  ,
\end{eqnarray}

\noindent in which $P$, $\theta$ and  $\tilde{Q}$ are given respectively by 
eqs.(\ref{A.7a}), (\ref{A.9a}), (\ref{A.9b}). The HFE (\ref{6}) is an
even function of $h$: ${\mathcal W}_1 (J, h, D; \beta) =
{\mathcal W}_1 (J, -h, D; \beta)$.
One is reminded that
${\mathcal W}_1^{stag} (J, h^{\prime}, D^{\prime}; \beta) =
{\mathcal W}_1 ( - J, h^{\prime}, D^{\prime}; \beta)$
and ${\mathcal W}_{Hub} (U, V, \mu; \beta) =
{\mathcal W}_1 ( J, h^{\prime}, D^{\prime}; \beta) 
- (J + h^{\prime} + D^{\prime})$, with $U = 2 D^{\prime}$,
$V = \frac{J}{2}$ and $\mu = 2J + h^{\prime} + D^{\prime}$.
 
From the simple expression (\ref{6}) one can calculate  the 
thermodynamic functions of the standard and the 
staggered models, as well of the extended Hubbard model in 
the ionic limit in the absence of an external magnetic 
field, for any finite value of $\beta$, that does not 
include the temperature $T=0$, that is,
$T \in (0, \infty)$.   It is a well known
fact that at $T=0$, $h=0$ and $D=0$, two eigenvalues of matrix 
{\bf U} are degenerated. Our results are valid in the limit of 
$T \rightarrow 0$. Due to the presence of the longitudinal 
magnetic field  in the HFE (\ref{6}) we can derive from it the 
$z$-component of the magnetization and the magnetic susceptibility 
per site for the ferromagnetic ($J<0$) and AF ($J>0$)
versions of the standard Hamiltonian (\ref{1})/ (\ref{5})
as well as the staggered  magnetization  and the staggered magnetic 
susceptibility\cite{physA2011} of the Hamiltonian (\ref{4}) for 
any value of the  exchange strength $J$. For $h=0$, 
the expression (\ref{6}) of the HFE ${\mathcal W}_1 (J, 0, D; \beta)$
coincides with the exact result presented in Ref.\cite{winder}. 

Our exact result permits us to study the plateaus of the 
standard and the staggered $z$ component of the 
magnetization per site of the ferro and the AF
models of the standard and the staggered versions, respectively,
of the one-dimensional spin-1 Ising model, with the single-ion
anisotropy term, in the presence of a longitudinal magnetic field 
for temperatures close to $T\sim 0$. 
The authors of Ref.\cite{krinsky} mentioned the discontinuity
of the magnetization and  the magnetic quadrupolar moment
of the ferromagnetic model at $T=0$, not going any further on this point. 
Aydiner and Aky\"uz \cite{aydiner} and
Chen {\it et al} \cite{chen} studied 
numerically these plateaus in the AF version 
of Hamiltonian (\ref{1})/(\ref{5}).
More recently, Mancini and Mancini\cite{mancini2008} solved 
(also numerically) the self consistent equations that 
yield the exact HFE of the AF $S=1$  Ising model
with a biquadratic nearest-neighbour term in the presence
of a longitudinal magnetic field.

\vspace{0.3cm}

In the rest of this paper we let $J= -1$ for the ferromagnetic
model and $J= 1$  for the AF model. The values of 
the parameters $D$, $h$ and $T$ are given in units of $|J|$, that is:
$\frac{D}{|J|}$, $\frac{h}{|J|}$ and $\frac{T}{|J|}$, respectively.
	
We restrict our presentation mainly to the 
behavior of two thermodynamic functions associated to 
the Hamiltonians (\ref{1})/(\ref{5}) and (\ref{4})
that helps us to understand the presence
of plateaus in the classical spin-1 model: 
the $z$-component of the magnetization\cite{meio}, 
${\mathcal M}_z  (J, h, D; \beta) \left(
 {\mathcal M}_z = - \frac{1}{2} \; \frac{\partial {\mathcal W}_1}{\partial h}
\right)$, of the standard model (\ref{1})/(\ref{5}),  and 
the staggered  $z$-component of the magnetization\cite{physA2011}, 
${\mathcal M}_z^{stag}  (J, h, D; \beta) \left(
{\mathcal M}_z^{stag} = 
\frac{1}{2} \left[ \langle S_2^z\rangle  - \langle S_2^z\rangle  \right] = 
 - \frac{1}{2} \; \frac{\partial {\mathcal W}_1^{stag}}{\partial h}
\right)$, of the staggered Hamiltonian  (\ref{4}); and 
the entropy  per site,
${\mathcal S}  (J, h, D; \beta) \left(
 {\mathcal S} =  \beta^2 \; \frac{\partial{\mathcal W}}
 {\partial \beta} \right)$, where ${\mathcal W} = 
{\mathcal W}_1 ({\mathcal W}_1^{stag})$ for the 
standard (staggered) model.
Those are studied at very low temperatures. The HFE's ${\mathcal W}_1$
and ${\mathcal W}_1^{stag}$ are  even functions of $h$,
 therefore ${\mathcal S}  (J, h, D; \beta)$ and 
the magnetization functions (${\mathcal M}_z  (J, h, D; \beta)$ 
and ${\mathcal M}_z^{stag}  (J, h, D; \beta)$) are even and 
odd functions of $h$, respectively.  For this reason 
we restrict ourselves to the case of $h\ge 0$. 

In Fig.\ref{fig_1} we present the phase diagram of the ferro
($J= -1$) and AF versions of Hamiltonian 
(\ref{1})/(\ref{5}) at $T=0$.  In its caption we describe the 
meaning of the phases in each diagram.
It is interesting to notice that the exchange coupling 
term $J$ in the ferromagnetic Hamiltonian (\ref{1})/(\ref{5})
favors the states with $z$ component of the spin $s_i^z = \mp 1$
and parallel neighboring spins.
The same term in the AF version of this Hamiltonian also 
favors the states with $s_i^z = \mp 1$ but for anti-parallel 
(N\'eel state) neighboring spins.  In the ferro and AF models the 
Zeeman term favors states with $s_i^z = \mp 1$ aligned
with the external magnetic field $h$, whereas the 
single-ion anisotropy term  shows two distinct 
behaviors: for $D<0$ states with
$s_i^z = \mp 1$ are favored, independently of their relative alignment;
and for $D>0$ states with  $s_i^z = 0$ are favored, 
the spin being perpendicular to the external magnetic 
field applied. 
For $D < 0$ all the terms in Hamiltonian (\ref{1})/(\ref{5})
force the neighboring spins to align to each other  and 
consequently to the external magnetic field at $T=0$. The ground state 
is a collective stable state  under small temperature fluctuations,
a point which will be made clear after  we discuss
the necessary temperature to excite the 
first excited state of the ferromagnetic chain. For $D > 0$ the 
effect of the crystal field $D$  competes with the exchange strength $J$ 
and the external magnetic $h$. In the near future we will verify that
in this case we need a smaller energy to break the alignment 
between neighboring spins and excite the first excited 
energy level of the  ferromagnetic chain.  

For $\frac{h}{|J|} \ge 0$
and only for $\frac{D}{|J|} \ge \frac{1}{2}$ the Fig.\ref{fig_1}a 
has two distinct phases. The line between the phases is:
$\frac{h}{|J|} =  \frac{D}{|J|} - \frac{1}{2}$. 
In the region  A  of Fig.\ref{fig_1}a, at $T=0$, 
the single-ion anisotropy term gives the main contribution 
to the ground state and the spins are perpendicular to the
longitudinal magnetic field. Our
phase diagram Fig.\ref{fig_1}a coincides with Fig.2 of 
Ref.\cite{krinsky} with $K=0$.

Fig.\ref{fig_1}b presents the phase diagram of the 
standard AF model  ($J=1$)  at $T=0$.  For $\frac{D}{|J|} <0$ 
and $\frac{h}{|J|} \ge 0$
there are two distinct phases that correspond to the 
competition between  the exchange coupling 
term $J$ and the Zeeman term.
One of them is the phase G that is 
the N\'eel state when $0 \le \frac{h}{|J|} < 1$; 
the exchange term gives the largest contribution to 
the ground state energy.  For $\frac{D}{|J|} > 0$ 
and $\frac{h}{|J|} \ge 0$ the
three terms in the AF Hamiltonian (\ref{1})/(\ref{5}) favor
distinct states at $T=0$. Their competition is responsible
for a richer phase diagram at $T=0$ for the AF model in this
region of parameters.  
On the other hand, this competition among the terms
of Hamiltonian (\ref{1})/(\ref{5}) makes the ground state of the
one-dimensional system less stable under small fluctuations of the
thermal energy, as we will verify in the discussion of the
first excited state of the AF chain.
The diagram of this 
model has two tricritical points: ${\mathcal P} (\frac{h}{|J|} = 1$
and $D=0$)  and  ${\mathcal Q} (\frac{h}{|J|} = \frac{1}{2}$
and $\frac{D}{|J|} = \frac{1}{2}$) in Fig.\ref{fig_1}b. For
$0< \frac{D}{|J|} < \frac{1}{2}$ and 
$\frac{h}{|J|} \ge 0$, the AF model has three
distinct phases at $T=0$, namely: the G (N\'eel) phase, 
the E phase (in which half of the spins in the chain are 
perpendicular to the external magnetic field), and the
B phase (in which all the spins are aligned with the magnetic field). 
We also have three phases for 
$\frac{D}{|J|} \ge \frac{1}{2}$, but in this region  of the phase
diagram  at $T=0$ the N\'eel state is not one of the 
possible ground states of the system;
there is an A phase instead, in which all spins
are perpendicular to the longitudinal magnetic field,
besides the phases B and E. The phase diagram 
Fig.\ref{fig_1}b agrees with Fig.2b of Ref.\cite{chen}
and  shows qualitative agreement with Fig.2
of Ref.\cite{aydiner}.

The transitions between the ground states of the standard AF model
happen for the following values of magnetic field $\frac{h}{|J|} \ge 0$
and the intervals of $\frac{D}{|J|}$:
 
\begin{subequations}

\begin{eqnarray}
\mbox{G} \rightleftharpoons \mbox{B}: & \frac{D}{|J|} <0& \mbox{and} 
     \hspace{0.7cm} \frac{h}{|J|} =1 , 
                               \label{7a} \\
\mbox{G} \rightleftharpoons \mbox{E}: & 
     0<\frac{D}{|J|} < \frac{1}{2} & 
     \mbox{and} \hspace{0.7cm} \frac{h}{|J|} =1 - \frac{D}{|J|} ,
                               \label{7b} \\
\mbox{A} \rightleftharpoons \mbox{E}: & \frac{D}{|J|} > \frac{1}{2} 
    & \mbox{and} \hspace{0.7cm} \frac{h}{|J|} = \frac{D}{|J|} ,
                               \label{7c} \\
\mbox{E} \rightleftharpoons \mbox{B}: & \frac{D}{|J|} >0 
       & \mbox{and} \hspace{0.7cm} \frac{h}{|J|} =1 + \frac{D}{|J|} .
                               \label{7d} 
\end{eqnarray}

\end{subequations}

The phase diagram of the staggered ferromagnetic ($J = -1$) 
Hamiltonian (\ref{4})  at $T=0$ is shown in Fig.\ref{fig_1}b
whereas the diagram of the staggered AF ($J=1$)  model at $T=0$ is 
depicted in Fig.\ref{fig_1}a.
 
\vspace{0.5cm}

To the best of our knowledge, there is no detailed
discussion in  the current literature regarding the 
presence of plateaus in the $z$-component of the magnetization 
in $1D$ ferromagnetic
spin models at very low temperatures,
although its discontinuity is mentioned in 
Ref.\cite{krinsky}. 
From the exact expressions of the $z$-component of the 
magnetizations of the ferro and AF spin-1 classical models,
valid in the thermodynamic limit,  
we verify that they have no discontinuity for $T\stackrel{>}{_{\sim}}0$.
Rather, there is a range of values of the magnetic field  
for which a continuous transition from one 
magnetization plateau to another takes place. Our result (\ref{6}) 
is equally valid for the ferro and AF models of the standard 
and staggered spin-1 Ising model with $(S_i^z)^2$ term in the
presence of a longitudinal magnetic field 
$T \stackrel{>}{_{\sim}} 0$.

Our result (\ref{6}) of the HFE of the ferro ($J= -1$)
and the AF ($J=1$) Hamiltonians (\ref{1})/(\ref{5}) are valid for 
$\frac{T}{|J|} > 0$. A natural way to describe  those models at
finite temperature is through their respective 
density matrix operator. From now on, we restrict our discussions 
to the condition $\frac{h}{|J|} \ge 0$.


\section{The ferromagnetic classical spin-1 model } \label{sec_3}

Let us consider the ferromagnetic model (\ref{1})/(\ref{5})
with $J= -1$. The ground  states of the chain at $T=0$ in the 
phases A, B and C in Fig.\ref{fig_1}a are 

\vspace{-0.5cm}

\begin{subequations}

\begin{eqnarray} 
|\Psi_0 \rangle_A &=& |0 \rangle_1 \otimes |0 \rangle_2 \otimes \cdots 
\otimes |0 \rangle_N ,   \label{8a}   \\
|\Psi_0 \rangle_B &=& |1 \rangle_1 \otimes |1 \rangle_2 \otimes \cdots 
\otimes |1 \rangle_N ,   \label{8b}   \\
|\Psi_0 \rangle_C &=& |-1 \rangle_1 \otimes |-1 \rangle_2 \otimes \cdots 
\otimes |-1 \rangle_N .   \label{8c}  
\end{eqnarray}

\end{subequations}

\noindent where $S_i^z |s \rangle_i = s |s \rangle_i$, $s = 0, \pm 1$
and $ i = 1, 2, \cdots, N$. Their respective ground state energy
are named $E_0^A$, $E_0^B$ and $E_0^C$.  The ground state (\ref{8c}) is 
presented since at $h=0$ its energy  
degenerates into  the energy of the state $|\Psi_0 \rangle_A$. 

The values of the ground state energies in units of $|J|$ are

\vspace{-0.5cm}

\begin{subequations}

\begin{eqnarray}
\frac{E_0^A}{|J|} & = & 0 ,   \label{9a} \\
\frac{E_0^B}{|J|} & = & N \left( -1 - \frac{2h}{|J|} + \frac{2D}{|J|} \right) ,
                              \label{9b}  \\
\frac{E_0^C}{|J|} & = & N \left( -1 + \frac{2h}{|J|} + \frac{2D}{|J|}  \right) ,
                              \label{9c} 
\end{eqnarray}

\end{subequations}

\noindent where $N$ is the number of sites in the chain.

The density matrix operator of the ferromagnetic Hamiltonian 
(\ref{1})/(\ref{5})   is  

\vspace{-0.5cm}

\begin{eqnarray}  \label{10}
&& \rho \left( -1, \frac{D}{|J|}, \frac{h}{|J|}; \beta \right) 
= \frac{e^{- \beta E_0^A}}{Z_1} \; |\Psi_0 \rangle_A \hspace{-0.2cm}  \phantom{...}_A\langle\Psi_0|
+ \frac{e^{- \beta E_0^B}}{Z_1} \; |\Psi_0 \rangle_B \hspace{-0.2cm}  \phantom{...}_B\langle\Psi_0|
  +   \nonumber   \\
&& 
\hspace{1.5cm} + \frac{e^{- \beta E_0^C}}{Z_1} \; |\Psi_0 \rangle_C \hspace{-0.2cm}  \phantom{...}_C\langle\Psi_0|
+ \frac{e^{- \beta E_1}}{Z_1} \times \sum_{l=1}^N  \; |\Phi_1 ^{(l)} \rangle \langle \Phi_1^{(l)}|
      \cdots  ,
\end{eqnarray}

\noindent where $Z_1$ is the partition function associated to the ferromagnetic
Hamiltonian (\ref{1})/(\ref{5}), $Z_1 = Tr[ e^{-\beta {\mathbf{H}}}]$, 
with $\beta = \frac{1}{kT}$.  $E_1$  is the first excited state of the chain 
and its degeneracy is equal to $N$ for all three ferromagnetic phases. The first excited 
state is obtained from the ground state vectors (\ref{8a})-(\ref{8c}) by 
flipping one of its  spin-1 in the chain. In the thermodynamic limit, the 
first excited state is highly degenerate as well the higher excited 
states\cite{alcaraz}. In principle, 
the contribution of all excited states of the chain should
be taken into account for non null temperatures. 
We have  no mathematically sound argument to 
affirm that, at low temperatures, the expansion (\ref{10}) 
should be cut after the contribution of the  first excited state. The exact 
expressions of the thermodynamic functions can be derived  from
result (\ref{6}) of the HFE (\ref{6}) of the HFE, and they show 
plateaus in the $z$-component of the magnetization of this model
at low temperature. The comparison between the exact result and a
proposed approximation expression of ${\mathcal M}_z$  should be 
able to say how good is the latter. 

It is common sense that the the least value of temperature
for which the contribution of the first 
excited state must be taken into account for 
the behavior of the thermodynamic functions
is determined by the factor $e^{- \frac{E_1 - E_0}{kT}}$. But in 
the present model the excited states have degeneracy at least of order
$N$, so all of them should contribute to the functions at 
non zero temperature.

In the following discussion we determine in each phase of the
diagram \ref{fig_1}a the value of temperature $T_{max}$ where   

\vspace{-0.5cm}

\begin{eqnarray}   \label{11}
e^{- \frac{E_1 - E_0}{k T_{max}}} \approx e^{-15}  \approx 3.06 \times 10^{-7} , 
\end{eqnarray}

\noindent where $E_0$  is the ground state energy of the phase
and $E_1$ is its first excited state. Then

\vspace{-0.5cm}

\begin{eqnarray}  \label{12}
\frac{kT_{max}}{|J|}  \approx \frac{E_1 - E_0}{15} .
\end{eqnarray}

We want to verify if the temperature $T_{max}$ at each ferromagnetic  phase
is such that, for temperatures lower than $T_{max}$, the dependence of the 
$z$-component of the magnetization on the external magnetic field 
 has a ``step-like" form. 

We have three distinct first excited states of the ferro classical
spin-1 model (\ref{1})/(\ref{5}). They depend on the values 
of the  parameters  $\frac{h}{|J|}$ and  $\frac{D}{|J|}$.  

There are two distinct  first excited states in phase B (see
Fig.\ref{fig_1}a): 

\hspace{1cm} 1.1)  for $\frac{h}{|J|} > 0 $ and $\frac{D}{|J|} \le -1 - \frac{h}{|J|}$  

\vspace{-0.5cm}

\begin{eqnarray}  \label{13}
|\Phi_1^{(l)} \rangle_B &=& |1 \rangle_1 \otimes |1 \rangle_2 \otimes \cdots 
 \otimes |-1 \rangle_l \otimes \cdots \otimes |1 \rangle_N ,
    \hspace{1cm} l = 1, 2, \cdots, N,
\end{eqnarray}

\noindent  and  

\vspace{-0.5cm}

\begin{eqnarray}   \label{14}
 \frac{E_1^B - E_0^B}{|J|}  = 4 \left( 1 + \frac{h}{|J|}  \right)  .
\end {eqnarray}

\noindent Replacing the result (\ref{14}) in eq.(\ref{12}), 
we obtain  

\vspace{-0.5cm}

\begin{eqnarray}  \label{15}
\frac{k T_{max}}{|J|} \approx \frac{4}{15} \left(1 + \frac{h}{|J|} \right) .
\end{eqnarray}

\noindent That gives the lowest temperature for which the first excited 
state is expected to contribute to the behavior of the thermodynamic functions.
Eq.(\ref{15}) shows that $T_{max}$ is
independent of the value of $\frac{D}{|J|}$. In this region of parameters
of the Hamiltonian (\ref{1})/(\ref{5}), the action of the crystal field
$D$ term is more relevant than the exchange coupling term $J$ and 
the Zeeman term. 

\vspace{0.3cm}

\hspace{1cm} 1.2)  for   $\;\; -1 - \frac{h}{|J|} \le\frac{D}{|J|} \le \frac{1}{2}$
 and $\frac{h}{|J|} > 0   \;\; \cup \;\;  \frac{D}{|J|} > \frac{1}{2} $ and 
$\frac{h}{|J|} \ge \frac{D}{|J|}  - \frac{1}{2}  $  

\vspace{-0.5cm}

\begin{eqnarray}  \label{16}
|\Phi_1^{(l)} \rangle_B &=& |1 \rangle_1 \otimes |1 \rangle_2 \otimes \cdots 
 \otimes |0 \rangle_l \otimes \cdots \otimes |1 \rangle_N ,
    \hspace{1cm} l = 1, 2, \cdots, N,
\end{eqnarray}

\noindent and  

\vspace{-0.5cm}

\begin{eqnarray}   \label{17}
 \frac{E_1^B - E_0^B}{|J|}  = 2 \left( 1 + \frac{h}{|J|} - \frac{D}{|J|}  \right)  .
\end {eqnarray}

For this set of parameters $\frac{h}{|J|}$  and $\frac{D}{|J|}$, 
eq.(\ref{12}) gives

\vspace{-0.5cm}

\begin{eqnarray}  \label{18}
\frac{k T_{max}}{|J|} \approx \frac{2}{15} \left( 1 + \frac{h}{|J|} - \frac{D}{|J|}  \right)  .
\end{eqnarray}

The results (\ref{15}) and (\ref{18}) belong to the same phase B
and their difference with respect to the first excited state 
come from the fact that they describe the behavior of the chain of spins-1
in distinct ranges of the parameters of the Hamiltonian: in case (1.1) (case (1.2))
the effect of $\frac{D}{|J|}$ is more (less) important than the 
exchange coupling  and the term Zeeman effect together.

\vspace{0.3cm}

\hspace{1cm}  2) the first excited state in phase A (see Fig.\ref{1}a):
$\frac{D}{|J|} \ge \frac{1}{2}$ and  $ 0 \le \frac{h}{|J|} \le \frac{D}{|J|} - \frac{1}{2}$  

\vspace{-0.5cm}

\begin{eqnarray}  \label{19}
|\Phi_1^{(l)} \rangle_A &=& |0 \rangle_1 \otimes |0 \rangle_2 \otimes \cdots 
\otimes |1 \rangle_l \otimes \cdots 
\otimes |0 \rangle_N ,   \hspace{1cm} l = 1, 2, \cdots, N ,
\end{eqnarray}  

\noindent and

\vspace{-0.5cm}

\begin{eqnarray}  \label{20}
  \frac{E_1^A - E_0^A}{|J|} = 2 \left( - \frac{h}{|J|} + \frac{D}{|J|} \right).
\end{eqnarray}

Replacing the variation of energy (\ref{20}) in eq.(\ref{12}), 
we have  

\vspace{-0.5cm}

\begin{eqnarray} \label{21}
   \frac{kT_{max}}{|J|} \approx \frac{2}{15} \left( - \frac{h}{|J|} 
     + \frac{D}{|J|}   \right)   .
\end{eqnarray}

In the present case the value of $T_{max}$ depends on the value
of the crystal field $D$ and  $T_{max}$ is a
decreasing function of $\frac{h}{|J|}$. The latter implies that  the ground state 
vector (\ref{8a}) is less stable than the vector (\ref{8b}) and (\ref{8c})
under an increasing external magnetic field. In phase A of 
Fig.\ref{fig_1}a we have a competition between the crystal field
$D$ term and the other two terms in the ferromagnetic 
Hamiltonian (\ref{1})/(\ref{5}). 

In Fig.\ref{fig_2} we plot the curves $\frac{kT_{max}}{|J|}$ as a 
function of $\frac{h}{|J|}$ for three values of $\frac{D}{|J|}$, covering
all the regions of $\frac{h}{|J|} \ge 0$ described in the discussion 
of the first excited states of the ferromagnetic classical 
spin-1 model (\ref{1})/(\ref{5}). For   $\frac{D}{|J|} = 2.5$ we 
have phase A  at $T=0$  and there is a competition
between the action of the single-ion anisotropy term 
and the two other terms in the Hamiltonian 
(\ref{1})/(\ref{5}). For  $0 \le \frac{h}{|J|} \le 2$ we have a negative slope in the 
curve $\frac{kT_{max}}{|J|}$. This negative slope means that 
the first excited state can be reached without increasing the temperature.

In Figs.\ref{fig_3}a and \ref{fig_3}b  we plot the exact curves 
(solid lines) of ${\mathcal M}_z \times \frac{h}{|J|}$ 
with $\frac{D}{|J|} = -2 \;\; \mbox{and} \;\; 2.5$ at 
$\frac{kT}{|J|} = 0.266\;\; \mbox{and} \;\;
0.0665$, respectively. In Fig.\ref{fig_3}a 
the $z$-component of the magnetization of the 
ferromagnetic model  begins
at null value with $h =0$ and reaches its saturated  value 
${\mathcal M}_z = 1$ for variations of the magnetic field $\Delta h$ so that 
$\frac{\Delta h}{|J|} \sim 7\times 10^{-4}$.  With $\frac{D}{|J|} = 2.5$
the function ${\mathcal M}_z$ in Fig.\ref{fig_3}b has a continuous
transition from ${\mathcal M}_z = 0$ to ${\mathcal M}_z = 1$ 
for $\frac{\Delta h}{|J|} \sim 4 \times 10^{-4}$. The maximum
value of the  entropy per site for these two cases is equal to 
$4.62 \times 10^{-3}$ at $h =0$ and $\frac{h}{|J|} = 2$. 
These are the values of the external magnetic field that happens 
the ferromagnetic phase transitions at $T = 0$ in diagram \ref{fig_1}a.

In Fig.\ref{fig_3}c we compare the graphs of the $z$-component 
of the magnetization of the model with $\frac{D}{|J|} = 2.5$ at
two different temperatures: $\frac{kT}{|J|} = 0.0665$ (dotted line) 
and $0.11$ (solid line). The latter value of temperature 
is almost twice the value of
$\frac{k T_{max}}{|J|}$ with $h=0$.
It is clear from Fig.\ref{fig_3}b and \ref{fig_3}c the presence 
of the two plateaus at ${\mathcal M}_z = 0$
and ${\mathcal M}_z = 1$ at those low temperatures. These plateaus 
of ${\mathcal M}_z$  satisfy the Oshikawa, Yamanaka and Affleck (OYA)
condition\cite{OYA}. We remind that this 
condition, when applied to the periodic Hamiltonian (\ref{1})/(\ref{5})
with spin-1, imposes that $p (1 - {\mathcal M}_z)$ be an integer,  
where $p$ is the spatial period of the ground
state. This is a necessary condition for the occurrence
of a plateau in the magnetization  curve of the one-dimensional
spin system\cite{chen, OYA}. The same condition is satisfied 
in Fig.\ref{fig_3}a.   In Fig.\ref{fig_3}b, the transition 
between the two plateaus happens at $\frac{h}{|J|} = 2$, the  
value of the magnetic field per unit of $|J|$ in which this 
ferromagnetic model suffers a transition between the phases
A and B at $\frac{D}{|J|} = 2.5$. Note that 
doubling the temperature per unit of $|J|$
does not change much the "step-like" form of the  plateaus 
in the curve of ${\mathcal M}_z$. We could say that in the 
standard ferromagnetic  model the plateaus are stiff.

From Figs.\ref{fig_3} we verify  
that outside the transition region $\frac{\Delta h}{|J|}$ where
the value of ${\mathcal M}_z$ suffers a finite transition, the values 
of this thermodynamic function correspond to their respective value of 
the ground state in the phase diagram at $T = 0$. We can use
a phenomenological approach to fit  ${\mathcal M}_z$ in the whole
interval of $\frac{h}{|J|}$ at $T \stackrel{<}{_{\sim}} T_{max}$.
We assume that the  contributions  to ${\mathcal M}_z$ in the
transition region of $\frac{h}{|J|}$  at 
$T \stackrel{<}{_{\sim}} T_{max}$ come only from the ground states
in the density matrix operator (\ref{10}). We have two distinct 
situations to discuss.

\vspace{0.3cm}
 	
{\bf Situation 1):}  $\frac{h}{|J|} \stackrel{>}{_{\sim}} 0$  and
$\frac{D}{|J|} \le \frac{1}{2}$.
In this approximation  we obtain for 
$\frac{h}{|J|} \stackrel{>}{_{\sim}} 0$:

\vspace{-0.5cm}

\begin{eqnarray}  \label{22}
{\mathcal M}_z (J, D, h; \beta) = 
\frac{e^{- \beta E_0^B} - e^{- \beta E_0^C}}{e^{- \beta E_0^B} + e^{- \beta E_0^C}}
\;\; \stackrel{=}{_{_{E_0^B < E_0^C }}}  \;\;
   \tanh ( 2 \beta N h ) .
\end{eqnarray}

\noindent  Since we consider $\frac{h}{|J|}$ to be non-zero and
positive we have $E_0^B < E_0^C$.  We do not have, a priori, a 
mathematical sound argument to affirm that the approximation 
(\ref{22}) gives the right picture of the $z$-component 
of the magnetization in the ferromagnetic model at low  
temperatures.

On the last term on the r.h.s. of result (\ref{22}) we have 
a product of two limit processes in the region of $\frac{h}{|J|} \sim 0$: 
$N \rightarrow \infty$ (thermodynamic limit) and  
$\frac{h}{|J|} \rightarrow 0$. The result (\ref{22}) depends
on the assumption that
   
\vspace{-0.5cm}

\begin{eqnarray}  \label{23}
\lim_{\stackrel{N \rightarrow \infty}{\frac{h}{|J|} \rightarrow 0}}  N \cdot \frac{h}{|J|} = 
                   l (h;\beta),
\end{eqnarray}
	
\noindent in which the function $l (h; \beta)$ is assumed to be 
finite. The simplest phenomenological function $l (h; \beta)$ is a 
linear function

\vspace{-0.5cm}

\begin{eqnarray}  \label{24}
 l (h; \beta) = a \cdot \frac{h}{|J|} ,
\end{eqnarray}

\noindent where we take $a$ as a constant.

\vspace{0.5cm}

{\bf Situation 2:}  $\frac{h}{|J|} \sim \frac{D}{|J|}- \frac{1}{2}$  and
$\frac{D}{|J|} \ge \frac{1}{2}$.
Assuming that for $T  \stackrel{<}{_{\sim}}  T_{max}$ only the
ground states of the ferromagnetic phases A and B  contribute 
to ${\mathcal M}_z$ , we derive the approximate expression
of this thermodynamic function ${\mathcal M}_z$ valid for 
$\frac{D}{|J|} \ge \frac{1}{2}$ at low temperature,

\vspace{-0.5cm}

\begin{subequations}

\begin{eqnarray}   \label{25a}
{\mathcal M}_z (J, D, h; \beta) \approx 
\frac{e^{- \beta N \varepsilon}}{1 + e^{- \beta N \varepsilon}} ,
\end{eqnarray}

\noindent where

\begin{eqnarray}  \label{25b}
\varepsilon  \equiv 2 |J| \left[ \frac{h}{|J|} - 
      \left( - \frac{1}{2} + \frac{D}{|J|} \right) \right]  .
\end{eqnarray}

\end{subequations}

Again  in the exponential functions on the r.h.s. of eq.(\ref{25a})
we need to calculate the product of two limits in the region 
$ \frac{h}{|J|} \sim  \left( - \frac{1}{2} + \frac{D}{|J|} \right)$:
$N \rightarrow \infty$ and
 $ \frac{h}{|J|} \rightarrow  \left( - \frac{1}{2} + \frac{D}{|J|} \right)$.
In analogy to  eqs. (\ref{23}) and (\ref{24}) we write

\begin{eqnarray}  \label{26}
 \lim_{\stackrel{N \rightarrow \infty}{\frac{h}{|J|} - \left( - \frac{1}{2} + \frac{D}{|J|} \right)  \rightarrow 0}}
 N \cdot \left[\frac{h}{|J|} -  \left( - \frac{1}{2} + \frac{D}{|J|} \right)  \right]
  =  l (h;\beta) .
\end{eqnarray}

In this case we also take the simplest linear function for
the phenomenological function $ l (h;\beta)$,

\vspace{-0.5cm}

\begin{eqnarray}  \label{27}
 l (h;\beta) \equiv a \cdot  
 \left[\frac{h}{|J|} - \left( - \frac{1}{2} + \frac{D}{|J|} \right) \right] ,
\end{eqnarray}

\noindent where $a$ is a constant.

In Table \ref{tab_1} we present the values of the parameter
$a$  for various values of $\frac{D}{|J|}$ at distinct low temperatures.
Some of the values of $a$ were used 
in the approximate curves 
of ${\mathcal M}_z$  in Figs.\ref{fig_3}a and \ref{fig_3}b.

\section{The antiferromagnetic classical spin-1 model } \label{sec_4}

Fig.\ref{fig_1}b  shows the six distinct phases at T = 0 for 
the AF spin-1 Ising model with a single-ion anisotropy term in the 
presence of a longitudinal external magnetic field.
With null crystal field ($D = 0$), the model has only three different 
phases at $T=0$.

The three terms in the AF ($J = +1$) Hamiltonian (\ref{1})/(\ref{5}) 
compete: the exchange coupling term $J$ favors
neighbor spins to align anti-parallel. The most stable configurations
of the AF ground state happens when the effect of two terms on each spin 
in the chain are in the same direction.

We present in appendix \ref{Apend_B} the ground 
state vectors and energies of the six phases of the 
AF Hamiltonian (\ref{1})/(\ref{5}) at $T=0$ and their 
respective energy difference between the first excited state
and its ground state. We also present in that appendix the relation
between $\frac{k T_{max}}{J}$ and the parameters $\frac{h}{J}$
and   $\frac{D}{J}$ such that the condition (\ref{11}) is valid
for the classical AF spin-1 model.

In Figs.\ref{fig_4} we plot the curves of $\frac{k  T_{max}}{J}$
for various values of $\frac{D}{J}$ that spans all  AF phases at 
$T = 0$ in the phase diagram Fig.\ref{fig_1}b. The 
fundamental difference between Figs. \ref{2} and \ref{fig_4}  
of the ferromagnetic and AF models, respectively, is that the
temperature required to excite the first excited 
ferromagnetic state never vanishes.

In Fig.\ref{fig_4}a with $\frac{D}{J} < 0$ we have $T_{max} = 0$
at $\frac{h}{J} = 1$. The line $\frac{h}{J} =1$ with $\frac{D}{J} < 0$
separates the phases B and G in diagram \ref{fig_1}b. The result
$T_{max} =0$ along this line means that the classical AF 
spin-1 model (\ref{1})/(\ref{5}) is gapless along the separation 
of these two phases.

We also have $T_{max} = 0 $ in Fig.\ref{fig_4}b with 
$\frac{D}{J} = \frac{1}{3}$ at $\frac{h}{J} = \frac{2}{3}$ and $\frac{4}{3}$.
These values of $\frac{h}{J}$ are on the lines that 
separate the phases $E \rightleftharpoons G$ and 
$B \rightleftharpoons E$ respectively. For the interval
$0 \le \frac{D}{J} \le \frac{1}{2}$ the AF model (\ref{1})/(\ref{5})
is gapless along the lines $\frac{h}{J} = 1 - \frac{D}{J}$ and 
$\frac{h}{J} = 1 + \frac{D}{J}$. At $\frac{h}{J} = 2.5$ and 
$\frac{h}{J} = 3.5$ with $\frac{D}{J} = 2.5$ we also have 
$T_{max} = 0$ in Fig.\ref{fig_4}b. These values of $\frac{h}{J}$
are on the lines that separate the phases $A \rightleftharpoons E$ and 
$B \rightleftharpoons E$ respectively. The AF model (\ref{1})/(\ref{5})
is gapless for $\frac{D}{J} \ge \frac{1}{2}$ along the lines 
$\frac{h}{J} = \frac{D}{J}$ and  $\frac{h}{J} = 1+ \frac{D}{J}$.
We partially summarize the contents of the curves in 
Figs.\ref{fig_4} by saying that the classical AF spin-1 Hamiltonian
(\ref{1})/(\ref{5}) is gapless along the lines that separate 
the phases in the diagram \ref{fig_1}b at $T=0$.

\vspace{0.5cm}

The presence of plateaus in the thermodynamic functions 
${\mathcal M}_z$  of the AF Hamiltonian (\ref{1}) has been 
reported previously by some 
authors\cite{aydiner,chen,mancini2008} for positive
values of $\frac{D}{J}$\cite{chen}.  In the following
we discuss the behavior of the thermodynamic functions
${\mathcal M}_z$ and ${\mathcal S}$  of the AF version
of Hamiltonian (\ref{1}), that is $J= 1$, by varying
the value of the crystal field  per unit of $J$,
$\frac{D}{J}$, to span part of the  phase diagram 
of the AF model (\ref{1}) at $T=0$ (see Fig.\ref{fig_1}b).

Here we draw a more detailed comparison between
our phase diagram in Fig.\ref{fig_1}b
with the phase diagram of Ref.\cite{chen}, 
in which only the case $\frac{D}{J} > 0$ is considered. Before we continue 
the comparison, we should notice that  $h_{Chen} = 2h$ and
$D_{Chen} = 2 D$, where $h$ and $D$ are our parameters 
in the Hamiltonian (\ref{5}). Their Fig.2b agrees 
with our diagram \ref{fig_1}b for $\frac{D}{J} > 0$, 
except for the fact that in their phase diagram the 
phase ${\mathcal M}_z = m = 0$  does not 
distinguish the phases A and G (see the caption of
our Fig.\ref{fig_1}b) that correspond to different 
ground states. In our phase diagram of the AF model 
we point out the presence of two tricritical points,
that is, the points: ${\mathcal P} (\frac{h}{J} = 1$
and $D=0$)  and  ${\mathcal Q} (\frac{h}{J} = \frac{1}{2}$
and $\frac{D}{J} = \frac{1}{2}$).

For $\frac{D}{J} <0$ and $\frac{h}{|J|} > 0$ we have at $T=0$ 
the phases B and G  in diagram \ref{fig_1}b for the standard
AF model. In this region of parameters of the Hamiltonian  
we have a  competition between the states favored by the 
exchange coupling term and those favored by the Zeeman 
term.

Figs.\ref{fig_5} show the thermodynamic functions
${\mathcal M}_z$ and ${\mathcal S}$ versus $\frac{h}{J}$ 
with $\frac{D}{J} = -2$ (long-dashed lines).
The $z$-component of the magnetization for $\frac{D}{|J|} <0$ 
has two plateaus, ${\mathcal M}_z = 0 $ and  
${\mathcal M}_z = 1$, at very low temperatures. They satisfy the OYA
condition\cite{OYA}. In this region of $\frac{D}{J}$, the 
transition between these two plateaus happens at $\frac{h}{J} =1$,
that is the value in which the transition between the
phases B and G at $T=0$  also occurs, for 
$\frac{D}{J} <0$. Since along any
vertical line  in this region of $\frac{D}{J}$ 
only two phases are crossed at $T=0$ (see Fig.\ref{fig_1}b), 
the AF spin-1 Ising model  (\ref{1}) has only two plateaus 
at very low temperatures for $\frac{D}{J} <0$. 
Fig.\ref{fig_5}a shows the curve
${\mathcal M}_z \times \frac{h}{J}$ at $\frac{T}{J} = 0.266$,
that is the same temperature 
used in plotting this function for the ferro model
in Fig.\ref{fig_3}a with $\frac{D}{J} = -2$. 
The curve  of ${\mathcal M}_z$  at $\frac{T}{J} = 0.266$ still
has little resemblance to a ``step-like" function. 
This behavior differs from that of the magnetization
of the standard ferromagnetic model (see Fig.\ref{fig_3}a). 
By comparing the behavior of the plateaus in the 
magnetization in the standard 
ferro and the AF models at $\frac{T}{J} = 0.266$,
we can say that the plateaus of the 
latter smear out even at low temperatures. 
This happens because although the neighbour spins
are aligned (anti-aligned) in phase B (G)
the exchange coupling term  and the Zeeman term
together with the crystal field $D$ 
compete in promoting opposing effects  on these spins.
Fig.\ref{fig_5}b presents the entropy per site with 
$\frac{D}{J} = -2$ at $\frac{T}{J} = 0.266$ (long-dashed line).
From this plot we verify that we cannot approximate the 
function ${\mathcal M}_z$ by 
taking into account only the contribution  of the two
ground states of the phases B and G as was done in the 
ferromagnetic version of the model. At $\frac{T}{J} =  0.266$,
the interval $\Delta h$ of the magnetic field for which 
the transition between the plateaus ${\mathcal M}_z =0$
and ${\mathcal M}_z =1$ occurs is such that  
$\frac{\Delta h}{J} \sim 1$. For temperatures of three orders 
of magnitude lower, the function ${\mathcal M}_z$  as a 
function of $\frac{h}{J}$ has a ``step-like" form and 
$\frac{\Delta h}{J} \sim 10^{-3}$.

Next we consider the AF model (\ref{1}) with $\frac{D}{|J|} = 2.5$.
In Fig.\ref{fig_5}a we plot  ${\mathcal M}_z$ versus $\frac{h}{J}$  
at $\frac{T}{J} = 0.065$ (solid line). 
We verify
that at this  temperature three plateaus still occur; namely, at
${\mathcal M}_z \in \{ 0, \frac{1}{2}, 1\}$, for which  the OYA 
condition is satisfied. We have three plateaus in this case because the 
 vertical line in the diagram \ref{fig_1}b,
localized at $\frac{D}{|J|} = 2.5$,  crosses three
phases  (B, E and G). The values 
$\frac{h}{J} = 2.5$ and $3.5$  where 
the transitions between the plateaus of  the $z$-component 
of the magnetization occur is the same 
as that of the transition between the AF phases
$\mbox{A} \rightleftharpoons \mbox{E}$ and 
$\mbox{E} \rightleftharpoons \mbox{B}$, respectively,
of the standard model  (\ref{1}) 
at $T=0$. Also in this case the curve of 
${\mathcal M}_z$  looses its "step-like" form  at 
$\frac{T}{J} = 0.266$. 
The width of the transition between the plateaus of 
${\mathcal M}_z$ is $\frac{\Delta h}{J} \sim 0.5$ . This width
reduces to $10^{ -3}$ for temperatures three orders of
magnitude lower than $0.0665$

To understand why the AF function ${\mathcal M}_z$ 
looses its ``step-like" form  at $\frac{T}{J} = 0.266$ and $0.665$
with $\frac{D}{J} =  -2$ and $2.5$ respectively, we plot 
in Fig.\ref{fig_5}b the function ${\mathcal S}$ as a function
of $\frac{h}{J}$ for these two set of parameters. The function 
${\mathcal S}$ is non-null around $\frac{h}{J} = 1$ with
$\frac{D}{J} = -2$ (long-dashed line)  and around the 
values $\frac{h}{J} = 2.5$ and $3.5$ (solid line). These
points in the phase diagram \ref{fig_1}b
are on the transitions line between 
the phases $\mbox{A} \rightleftharpoons \mbox{E}$ and 
$\mbox{E} \rightleftharpoons \mbox{B}$ respectively,
of the standard model  (\ref{1})/(\ref{5}) with $\frac{D}{J} = 2.5$
at $T=0$. Although the AF models are presented at very low 
temperature, the excited states of the AF model already 
contribute to the thermodynamic function ${\mathcal M}_z$  around the
values of $\frac{h}{J}$ that correspond to the lines that 
separate the AF phases in Fig.\ref{fig_1}b.

The curve ${\mathcal M}_z \times \frac{h}{J}$ with 
$\frac{D}{J} = \frac{1}{3}$ and $2.5$ are very similar
at low temperatures. For $\frac{D}{J} = 0.25$, 
the transition from the plateaus ${\mathcal M}_z = 0$ to 
${\mathcal M}_z = \frac{1}{2}$ corresponds, in the 
phase diagram \ref{fig_1}b, to the transition $\mbox{G} \rightleftharpoons \mbox{E}$,
whereas for $\frac{D}{J} = 2.5$ (see Fig.\ref{5}a)  the same transition
corresponds to the phase crossover
$\mbox{A} \rightleftharpoons \mbox{E}$. The magnetization 
functions are almost insensible  if  the initial configuration
 of the spins is either a N\'eel state  or 
a state with all the spins perpendicular to the
longitudinal magnetic field. Comparing Figs.\ref{fig_6}a
and \ref{fig_6}b, we see that  the specific heat per site,
${\mathcal C}  (J, h, D; \beta) \left(
 {\mathcal C} = - \beta^2 \; \frac{\partial^2 [\beta{\mathcal W}_1]}
 {\partial \beta^2} \right)$ distinguishes
the transitions 
$\mbox{A} \rightleftharpoons \mbox{E}$ and $\mbox{G} \rightleftharpoons \mbox{E}$
at very low temperatures. In Fig.\ref{fig_6}a the function ${\mathcal C}$
is symmetric around $\frac{h}{|J|} = 0.75$.

From Fig. \ref{fig_5}a we could say that the plateaus of 
${\mathcal M}_z$ in the AF model  (\ref{1})/(\ref{5}) are soft, in 
the sense that they are already smeared out at $\frac{T}{J}  = 0.26$
and $0.0665$ with $\frac{D}{J}  = -2$ and $2.5$ respectively.

\vspace{0.5cm}

Since ${\mathcal W}_1^{stag} (J, h^{\prime}, D^{\prime}; \beta) =
{\mathcal W}_1 ( - J, h^{\prime}, D^{\prime}; \beta)$, we obtain

\begin{eqnarray}  \label{28}
{\mathcal S}^{stag}  (J, h, D; \beta) = {\mathcal S}  ( - J, h, D; \beta)
\end{eqnarray}

\noindent and

\begin{eqnarray}\label{29}
{\mathcal M}_z^{stag} (J, h, D; \beta) = {\mathcal M}_z ( - J, h, D; \beta) .
\end{eqnarray}

The plateaus in the $z$-component  of the magnetization
of the standard ferromagnetic (AF) model (\ref{1}) appear
in the $z$-component of the staggered magnetization
of the staggered AF (ferromagnetic) model (\ref{5}). 
This fact implies that the plateaus in the staggered 
magnetization of Hamiltonian (\ref{4}) satisfy
the OYA condition, but for this thermodynamic function
the AF model has only two plateaus for $\frac{D}{|J|} >0$.

The behavior of the curves of ${\mathcal S}^{stag}$ versus
$\frac{h}{|J|}$  of the staggered ferro  and AF models 
is identical to the curves of the entropy per site 
of the standard AF and ferromagnetic models (\ref{1})/(\ref{5}) 
respectively. It is very simple to understand this 
situation once the phase diagram at $T=0$ of the 
staggered ferromagnetic and AF models are depicted by 
Figs.\ref{fig_1}b and \ref{fig_1}a respectively.


\section{Conclusions} \label{sec_5}

The  density matrix method\cite{baxter}
and the $\beta$-expansion of the HFE\cite{chain_m} 
have been applied to the obtainment of
the exact expressions of the HFE's of the one-dimensional 
standard\cite{isingSz2} and the staggered\cite{physA2011}
spin-1 Ising models with $(S_z)^2$ term in the presence of
a longitudinal magnetic field. The analytic 
expressions of these functions  have been written  in terms
of exponentials of the parameters of the Hamiltonians 
(\ref{4}) and (\ref{1})/(\ref{5}). Our results are valid for their 
respective ferromagnetic ($J<0$) and  AF ($J>0$) models
in the interval $T \in (0, \infty)$, extending the results 
of Ref.\cite{krinsky} to the AF 
models. These results do not coincide with the results of Ref.\cite{litaiff}. 
We present the phase diagram of the standard and the 
staggered spin-models (\ref{1})/(\ref{5}) and (\ref{4}), 
respectively, at $T=0$. Our result also gives the 
exact HFE of the one-dimensional extended Hubbard model 
in the ionic limit but in the absence of an external
magnetic field.

We have studied  the behavior of the $z$-component  of the 
magnetization ${\mathcal M}_z$ and the entropy
 ${\mathcal S}$, both per site, of the 
ferromagnetic and AF models of Hamiltonian 
(\ref{1})/(\ref{5}) at  low temperatures
and their respective staggered  versions.

We have presented the two plateaus of ${\mathcal M}_z$ at
low temperatures of the standard ferromagnetic 
model for $\frac{D}{|J|} > \frac{1}{2}$ and 
show that the value in which the transition
between plateaus at  low temperatures occurs is the 
same as that of diagram \ref{fig_1}a.

For the AF model (\ref{1})/(\ref{2}) we show  that the number
of plateaus  of the $z$-component of the magnetization
at very low temperatures depends on the number of phases 
of the model at $T=0$ for a given value of  $\frac{D}{|J|}$. 
Our results for the AF model  with $\frac{D}{|J|} > 0$
agree with previous results in the 
literature\cite{aydiner,mancini2008,chen}.

The ferromagnetic classical spin-1 model (\ref{1})/(\ref{5})
has has highly degenerate excited states, but we showed that 
for temperatures $T\stackrel{<}{_{\sim}} T_{max}$ the function
${\mathcal M}_z$ can be approximated by the contribution
of only two ground states. This fact is not true for the
AF model because it is gapless for values of $\frac{h}{J}$
and $\frac{D}{J}$ along the line separating
the AF phases in the diagram  \ref{fig_1}b at $T=0$.

The AF specific heat at very low 
temperatures distinguishes the phase transitions
$\mbox{E} \rightleftharpoons \mbox{G}$ (at $\frac{h}{J} = 0.75$)
and  $\mbox{E} \rightleftharpoons \mbox{B}$ (at $\frac{h}{J} = 1.25$) 
at  $\frac{D}{|J|} = 0.25$  and $2.5$, respectively,
in Figs.\ref{fig_6}.

By comparing the plots of the ${\mathcal M}_z$
as a function of $\frac{h}{J}$  at 
$\frac{T}{|J|} = 0.266$ and $0.0665$ to
$\frac{D}{J} = -2$ and $ 2.5$ respectively,
we verify that the plateaus of this function of the standard  
ferro and AF  models can be called stiff and soft, 
respectively. 

All the plateaus in the $z$-component of the magnetization
of the ferro and AF models of the  standard and staggered
version satisfy the OYA condition.

As a final comment we should say that the presence of 
plateaus in the function ${\mathcal M}_z$ in the classical 
spin-1 Ising  model with single-ion anisotropy 
term in the presence of a 
longitudinal magnetic field comes from the stability  at low 
temperature of the ground state vector under the action of
increasing the norm of the external magnetic field and the 
temperature.

\vspace{1cm}

S.M.S. (Fellowship CNPq, Brazil, Proc.No.: 303562/2009-9) thanks  CNPq  
and FAPEMIG for the partial  financial support. M.T.T.  also thanks  
 FAPEMIG.  M.T.T. thanks J. Flor\^encio Jr and J.F. Stilck for enlightening 
discussions.


\appendix

\setcounter         {equation}{0}
\def\theequation{A.\arabic{equation}}

\section{Calculation of the roots of the third degree \hfill \\
equation of the spin-1 classical model.}  \label{Apend_A}

Krinsky and Furman  applied the transfer matrix 
method\cite{kramers1,kramers2,baxter} to calculate the HFE of the
classical spin-1 model in Ref.\cite{krinsky}. The roots coming from 
the cubic equations have to be real, but they write them in terms 
of complex quantities. When we handle those roots numerically these
complex quantities may not disappear. In this appendix we recalculate their
results for Hamiltonian (\ref{1})/(\ref{5}),
showing explicitly that the roots of the cubic equation are real. 

\vspace{0.3cm}

Following Ref.\cite{baxter} we obtain that the partition function
${\mathcal Z}_1 (J, h, D; \beta)$ of the spin-1 Hamiltonian
(\ref{1})/(\ref{5}) is equal to

\begin{eqnarray} \label{A.1}
{\mathcal Z}_1 (J, h, D; \beta) = Tr[{\bf U}^N],
\end{eqnarray}
 
\noindent where $N$ is the number of sites in the periodic chain
and the matrix {\bf U} for the symmetric Hamiltonian (\ref{5}) is

\begin{eqnarray}  \label{A.2}
{\bf U} (J, h, D; \beta) =
\left[
   \begin{array}{c c c}
  e^{- \beta(J + 2h+2D)} & e^{- \beta( h + D)} & e^{- \beta( -J + 2D)} \\
  e^{- \beta( h + D)} & 1 & e^{- \beta( -h + D)}  \\
  e^{- \beta( -J + 2D)} & e^{- \beta( -h + D)} &  e^{- \beta(J - 2h+2D)}
   \end{array}
\right]  .
\end{eqnarray}

\noindent The matrix ${\bf U} (J, h, D; \beta)$ 
is hermitian for any value of $J$, $h$, $D$ and $\beta$. 
Its three eigenvalues $\lambda_i$, $i= 1, 2$ and $3$, are 
real. The matrices ${\bf U}$ (see eq.(\ref{A.2})) and ${\bf T}$ 
(in Ref.\cite{krinsky}) differ by a rearrangement of 
lines, only.

In terms of the eigenvalues of {\bf U}, the partition function
(\ref{A.1}) becomes:

\begin{eqnarray}  \label{A.3}
{\mathcal Z}_1 (J, h, D; \beta) = (\lambda_1)^N \left[
1 + \left( \frac{\lambda_2}{\lambda_1}\right)^N 
 + \left( \frac{\lambda_3}{\lambda_1}\right)^N 
\right]  ,
\end{eqnarray}

\noindent in which $\lambda_1$ is assumed to be 
the eigenvalue of matrix {\bf U} with the largest modulus.

In the thermodynamic limit ($N \rightarrow \infty$), the partition 
function (\ref{A.1}) of the model is

\begin{eqnarray}  \label{A.4}
{\mathcal Z}_1 (J, h, D; \beta) = (\lambda_1)^N  ,
\end{eqnarray}

\noindent for non-degenerate eigenvalues of {\bf U}.
The HFE of the model is

\begin{eqnarray}   \label{A.5}
{\mathcal W}_1 (J, h, D; \beta) = - \frac{1}{\beta}
\; \; \ln[\lambda_1 (J, h, D; \beta)].
\end{eqnarray}

The eigenvalues $\lambda_i$, $i= 1, 2, 3$, are roots of the cubic
equation

\begin{eqnarray}  \label{A.6}
- \lambda^3 + P \lambda^2 + Q \lambda + R = 0  ,
\end{eqnarray}

\noindent where 

\begin{subequations}

\begin{eqnarray}   
P  &=& 1 + 2 e^{-\beta (J+2D)} \cosh(2 \beta h) = tr[{\bf U}] 
                             \label{A.7a} \\
Q &=& 4 e^{-2 \beta D} \;\;  e^{- \frac{\beta J}{2}}
\cosh(2 \beta h) \sinh\left(\frac{\beta J}{2}\right)  
+  2  e^{-4 \beta D}  \;\; \sinh(2 \beta J),  \label{A.7b} \\
R &=& -8 e^{-4 \beta D} \left[\sinh\left(\frac{\beta J}{2}\right)\right]^2  \;\;
\sinh(\beta J)  .   \label{A.7c}
\end{eqnarray}

\end{subequations}

The roots of the cubic equation  (\ref{A.6})  are well
known\cite{schaum}

\begin{subequations}

\begin{eqnarray}
s_1 &=& 2  \sqrt{- \tilde{Q}} \;\; \cos\left(\frac{\theta}{3}\right) + \frac{P}{3} ,
                    \label{A.8a}  \\
s_2 &=& 2  \sqrt{- \tilde{Q}} \;\;  \cos\left(\frac{\theta}{3} + \frac{2\pi}{3}\right) + \frac{P}{3} ,
                    \label{A.8b}  \\ 
s_3 &=& 2  \sqrt{- \tilde{Q}}  \;\;  \cos\left(\frac{\theta}{3} - \frac{2\pi}{3}\right) + \frac{P}{3} ,
                    \label{A.8c} 
\end{eqnarray}

\end{subequations}

\begin{subequations}

\noindent where

\begin{eqnarray}  \label{A.9a}
\cos(\theta)  = \frac{\tilde{R}}{\sqrt{ (- \tilde{Q})^3}}    
\end{eqnarray}

\noindent with

\begin{eqnarray}   \label{A.9b}
\tilde{Q} = - \frac{3 Q + P^2}{9} 
\end{eqnarray}

\noindent and 

\begin{eqnarray}   \label{A.9c}
\tilde{R} = \frac{ 9 Q P + 27 R + 2 P^3}{54}  .
\end{eqnarray}

\end{subequations}

Our previous result $s_1$ does not agree with the expression
of $\lambda_{max}$ of Ref.\cite{litaiff};  we believe that 
there are some misprints in their eqs.(18)-(24).


\setcounter         {equation}{0}
\def\theequation{B.\arabic{equation}}

\section{The ground states of the classical AF 
spin-1 and the calculation of $T_{max}$ }  \label{Apend_B}

For the AF version of the Hamiltonian (\ref{1})/(\ref{5}) we 
assume that the chain has a even number of 
sites. Let $N = 2M$, in which $M$ is a positive integer.
In the thermodynamic limit ($N \rightarrow \infty$) we have 
$M \rightarrow \infty$. 

The ground state vectors at each phase in the diagram \ref{fig_1}b
at $T=0$ are

\vspace{-0.5cm}

\begin{subequations}

\begin{eqnarray}
|\Psi_0 \rangle_A & =& |0\rangle_1 \otimes  |0\rangle_2 \otimes 
\cdots   \otimes  |0\rangle_{2M},   \label{B.1a} \\
|\Psi_0 \rangle_B & =& |1\rangle_1 \otimes  |1\rangle_2 \otimes 
\cdots   \otimes  |1\rangle_{2M},   \label{B.1b} \\
|\Psi_0 \rangle_C & =& |-1\rangle_1 \otimes  |-1\rangle_2 \otimes 
\cdots   \otimes  |-1\rangle_V,   \label{B.1c} \\
|\Psi_0 \rangle_E & =& |0\rangle_1 \otimes  |1\rangle_2 \otimes 
|0\rangle_3 \otimes  |1\rangle_4 \otimes
\cdots   \otimes  |0\rangle_{2M-1} |1\rangle_{2M},   \label{B.1d} \\
|\Psi_0 \rangle_F & =& |0\rangle_1 \otimes  |-1\rangle_2 \otimes 
|0\rangle_3 \otimes  |-1\rangle_4 \otimes
\cdots   \otimes  |0\rangle_{2M-1} |-1\rangle_{2M},   \label{B.1e} \\
|\Psi_0 \rangle_G & =& |1\rangle_1 \otimes  |-1\rangle_2 \otimes 
|1\rangle_3 \otimes  |-1\rangle_4 \otimes
\cdots   \otimes  |1\rangle_{2M-1} |-1\rangle_{2M}.   \label{B.1f} 
\end{eqnarray}

\end{subequations}

\noindent One is reminded that $S_i^z |s\rangle_i = s |s\rangle_i$, 
$s = 0, \pm 1$ and $i = 1, 2, \cdots 2M$. 

We are interested in the discussion of the thermodynamic function
${\mathcal M}_z$ at low temperatures. This function has an even  dependence
on $\frac{h}{J}$; then we restrict our discussion to  $\frac{h}{J} \ge 0$.
In the following we do not mention the quantum behavior of the phases 
C and F at low temperature.

The value of the ground state energy of the phases A, B, E and
G are, respectively,

\vspace{-0.5cm}

\begin{subequations}

\begin{eqnarray}
\frac{E_0^A}{J}  & = & 0,   \label{B.2a}  \\
\frac{E_0^B}{J}  & = & N \left( 1+ \frac{2D}{J} - \frac{2h}{J} \right) ,   \label{B.2b}  \\
\frac{E_0^E}{J}  & = & N \left( \frac{D}{J} - \frac{h}{J} \right) ,   \label{B.2c}  \\
\frac{E_0^G}{J}  & = & N \left( -1+ \frac{2D}{J} \right) .   \label{B.2d}  
\end{eqnarray}

\end{subequations}

We have a much more complex distribution of first excited states
$E_1$  in the AF spin-1 Ising model  (\ref{1})/(\ref{5}) than in the
ferromagnetic version of the model.

Let us present the difference between the first excited   
and the ground states of each phase A, B, E and
G in diagram \ref{fig_1}b.

\vspace{0.3cm}

\noindent  1) Phase A: $\frac{D}{J} \ge \frac{1}{2}$ and 
$\frac{h}{J} \le \frac{D}{J}$. We have

\begin{eqnarray}  \label{B.3}
 \frac{E_1 - E_0^A}{J} = 2 \left( - \frac{h}{J} +  \frac{D}{J} \right) ,
\end{eqnarray}

\noindent which is substituted in eq.(\ref{12}) to give

\vspace{-0.5cm}

\begin{eqnarray}  \label{B.4}
\frac{k T_{max}}{J} \approx \frac{2}{15} \left( - \frac{h}{J} +   
\frac{D}{J} \right) .
\end{eqnarray}

\noindent  2) Phase B: we have two distinct first excited states

\vspace{0.3cm}

\noindent  2.1)  $\frac{D}{J} \le 0$ and  $\frac{h}{J} \ge 1 - \frac{D}{J} 
\;\;  \cup \;\; \frac{D}{J} \ge 0$ and  $\frac{h}{J} \ge 1 + \frac{D}{J} $. 
Then

\begin{eqnarray}  \label{B.5}
\frac{E_1 - E_0^B}{J}  = 2 \left( -1 + \frac{h}{J} - \frac{D}{J} \right)
\end{eqnarray}

\noindent and eq.(\ref{12}) gives  

\vspace{-0.5cm}

\begin{eqnarray}  \label{B.6}
\frac{k T_{max}}{J}  \approx  \frac{2}{15} \left( -1 + \frac{h}{J} - \frac{D}{J} \right) .
\end{eqnarray}

\noindent  2.2)  $\frac{D}{J} \le 0$ and  $1 \le \frac{h}{J} \le 1 - \frac{D}{J}$. 
Then

\begin{eqnarray}  \label{B.7}
\frac{E_1 - E_0^B}{J}  = 4 \left( -1 + \frac{h}{J} \right) ,
\end{eqnarray}

\noindent and eq.(\ref{12}) becomes  

\begin{eqnarray}  \label{B.8}
\frac{k T_{max}}{J}  \approx  \frac{4}{15} \left( -1 + \frac{h}{J} \right) .
\end{eqnarray}

\vspace{0.3cm}

\noindent 3) Phase E: in this phase we have three
distinct  first excited states.   

\vspace{0.3cm}

\noindent  3.1) $0 \le \frac{D}{J} \le \frac{1}{2}$  and 
$1 - \frac{D}{J}  \le \frac{h}{J} \le 1$. 
We have  

\begin{eqnarray}  \label{B.9}
\frac{E_1 - E_0^E}{J}  = 2 \left( -1 + \frac{h}{J} +  \frac{D}{J} \right) ,
\end{eqnarray}

\noindent with eq.(\ref{12}) giving   

\vspace{-0.5cm}

\begin{eqnarray}  \label{B.10}
\frac{k T_{max}}{J}  \approx \frac{2}{15} \left( -1 + \frac{h}{J} +  \frac{D}{J} \right) ,
\end{eqnarray}

\noindent 3.2)   $0 \le \frac{D}{J} \le \frac{1}{2}$  and 
$1 \le \frac{h}{J} \le 1 + \frac{D}{J}  \;\;  \cup \;\;
 \frac{D}{J} \ge \frac{1}{2}$  and 
$\frac{1}{2} + \frac{D}{J} \le \frac{h}{J} \le 1 + \frac{D}{J} $. We have

\begin{eqnarray}  \label{B.11}
\frac{E_1 - E_0^E}{J}  = 2 \left( 1 - \frac{h}{J} +  \frac{D}{J} \right) ,
\end{eqnarray}

\noindent which gives

\vspace{-0.5cm}

\begin{eqnarray}  \label{B.12}
\frac{k T_{max}}{J}  \approx \frac{2}{15} \left( 1 - \frac{h}{J} +  \frac{D}{J} \right) .
\end{eqnarray}

\noindent 3.3) $ \frac{1}{2} \le \frac{D}{J} \le 1$  and 
$1 \le \frac{h}{J} \le \frac{1}{2} + \frac{D}{J}  \;\;  \cup \;\;
 \frac{D}{J} \ge 1$  and 
$ \frac{D}{J} \le \frac{h}{J} \le \frac{1}{2} + \frac{D}{J} $. We have  

\begin{eqnarray}  \label{B.13}
\frac{E_1 - E_0^E}{J}  = 2 \left( \frac{h}{J} -  \frac{D}{J} \right) ,
\end{eqnarray}

\noindent that substituted in eq.(\ref{12}) gives   

\begin{eqnarray}  \label{B.14}
\frac{k T_{max}}{J}  \approx \frac{2}{15} \left( \frac{h}{J} -  \frac{D}{J} \right) .
\end{eqnarray}

\vspace{0.3cm}

\noindent  4) Phase G: in this phase we have two
distinct  first excited states. 

\vspace{0.3cm}

\noindent 4.1)  $ \frac{D}{J} \le -1$  and 
$0 \le \frac{h}{J} \le 1 \;\;  \cup \;\;
-1 \le \frac{D}{J} \le 0$  and 
$ \frac{h}{J} \le 1 + \frac{D}{J} $. We have  

\begin{eqnarray}  \label{B.15}
\frac{E_1 - E_0^G}{J}  = 4 \left(1 - \frac{h}{J} \right) .
\end{eqnarray}  

From eq.(\ref{12}) we obtain

\begin{eqnarray}  \label{B.16}
\frac{k T_{max}}{J}  \approx \frac{4}{15} \left(1 - \frac{h}{J} \right) .
\end{eqnarray} 

\noindent  4.2)  $ -1 \le \frac{D}{J} \le 0$  and 
$\frac{h}{J} \le 1 + \frac{D}{J} \;\;  \cup \;\;
0 \le \frac{D}{J} \le \frac{1}{2}$  and 
$  \frac{1}{2} \le \frac{h}{J} \le 1 - \frac{D}{J} 
 \;\;  \cup \;\; 0 \le \frac{D}{J} \le \frac{1}{2}$ and 
 $0 \le \frac{h}{J} \le \frac{1}{2}$. We have   

\begin{eqnarray}  \label{B.17}
\frac{E_1 - E_0^G}{J}  = 2 \left(1 - \frac{h}{J} - \frac{D}{J} \right) ,
\end{eqnarray}

\noindent with eq.(\ref{12}) given 

\vspace{-0.5cm}

\begin{eqnarray}  \label{B.18}
\frac{k T_{max}}{J}  \approx \frac{2}{15} \left(1 - \frac{h}{J} - \frac{D}{J} \right) .
\end{eqnarray}



\setcounter {figure}{0}


\begin{figure} 
\begin{center}
\includegraphics[width=7cm,height= 7cm,angle= 0]{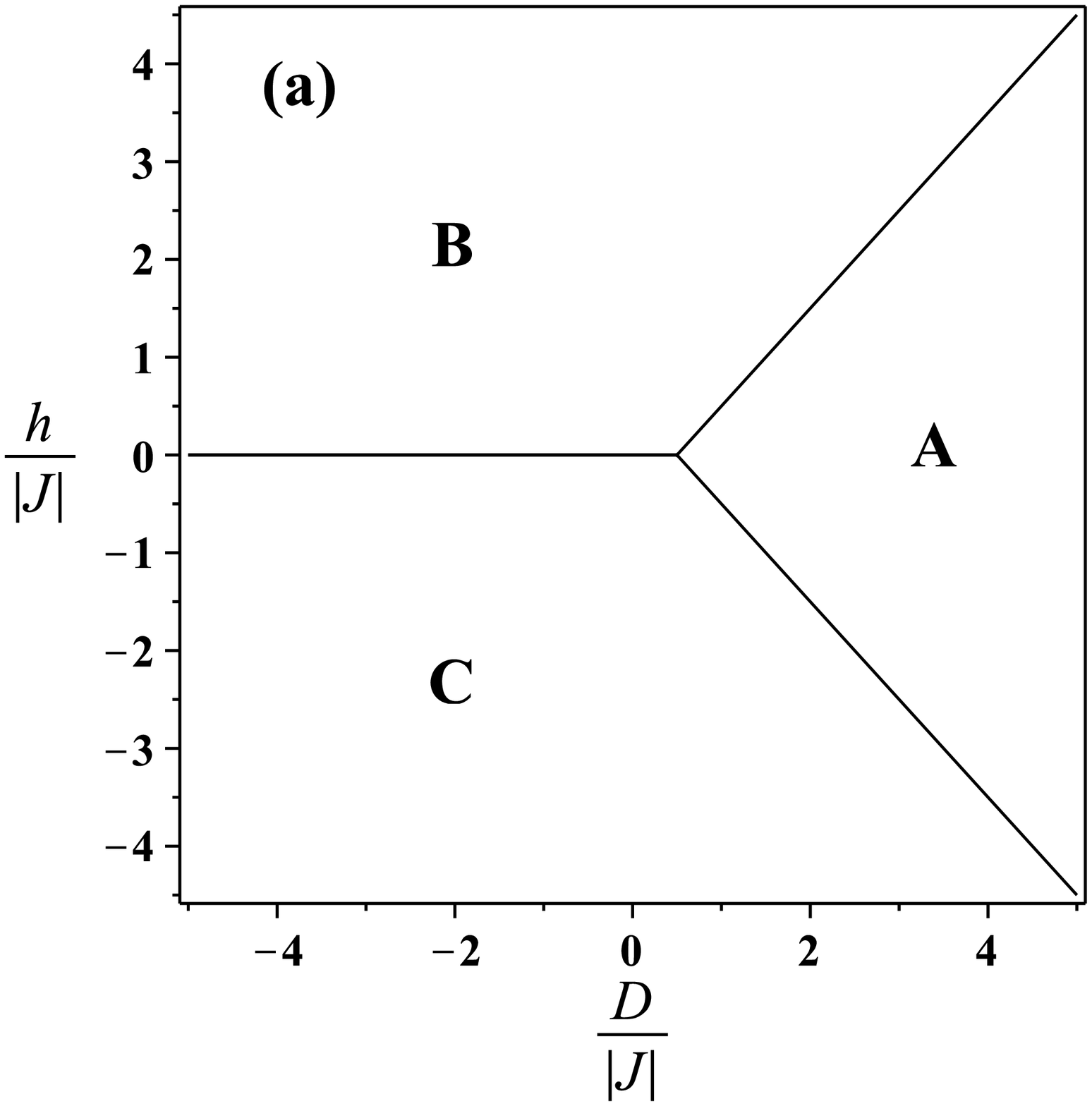}
\hspace{0.5cm}  
\includegraphics[width=7cm,height= 7cm,angle= 0]{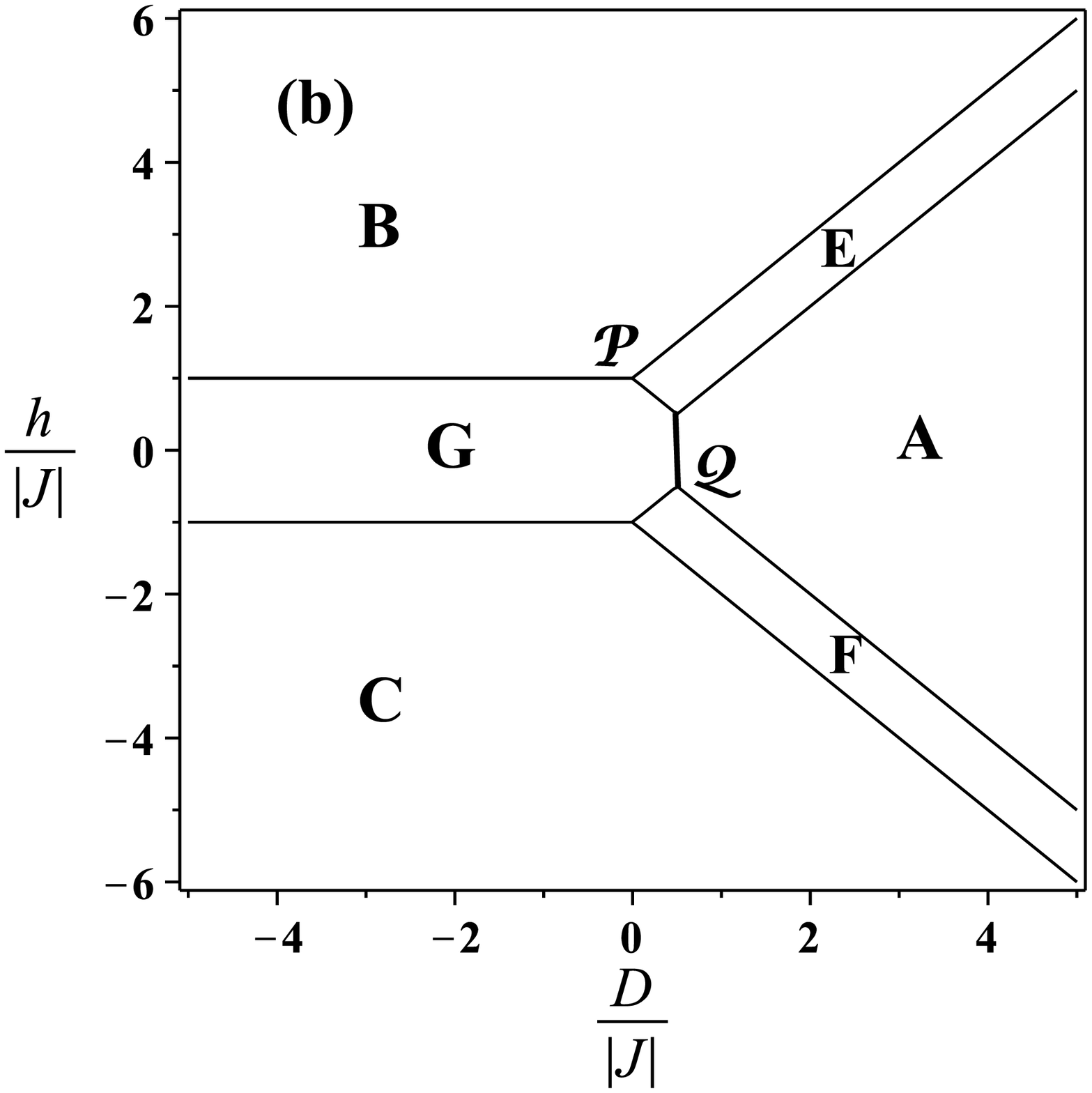}  
\vspace{-0.2cm}
\caption{The phase diagrams of the ferromagnetic  ($J= -1$)  and 
the anti-ferromagnetic ($J= 1$) models  (\ref{1})/(\ref{2})  at $T=0$.
({\it a}) The ferromagnetic phases represent  the following configurations 
of the $z$-components  of the neighbour spins: 
$\mbox{A} \rightarrow (0,0)$, $\mbox{B} \rightarrow (1,1)$ and $\mbox{C} \rightarrow (-1, -1)$.
({\it b})  The AF phases include the configurations A,
B and C that are found in the ferromagnetic diagram plus the phases:
 $\mbox{E}  \rightarrow (1,0)$, $\mbox{F} \rightarrow (0,-1)$ and $\mbox{G} \rightarrow (-1,1)$.
This diagram has also two tricritical points: ${\mathcal P} (\frac{h}{|J|} = 1$
and $D=0$)  and  ${\mathcal Q} (\frac{h}{|J|} = \frac{1}{2}$
and $\frac{D}{|J|} = \frac{1}{2}$). The phase diagram of the staggered
ferromagnetic and the AF models of the  Hamiltonian (\ref{4}),  
at $T=0$, are given by the figures ($b$)  and ($a$) respectively.
}  \label{fig_1}
\end{center}  
\end{figure}


\begin{figure}
\begin{center}
\includegraphics[width=7cm,height= 7cm,angle= 0]{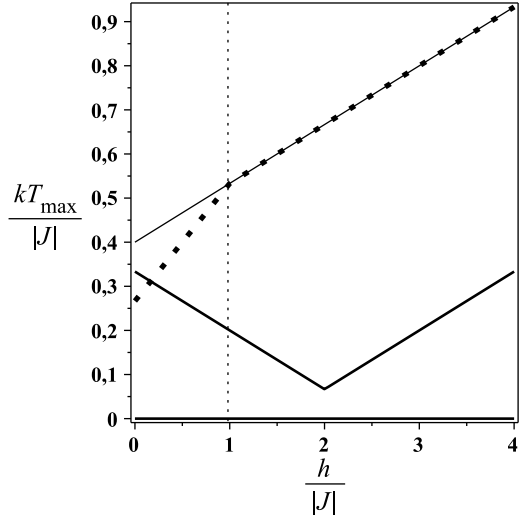} 
\vspace{-0.5cm}
\caption{
The curves of $\frac{k T_{max}}{|J|}$ for the three distinct
first excited states of the ferromagnetic model.  The dotted line
has $\frac{D}{|J|} = -2$, the solid line has $\frac{D}{|J|} = -\frac{1}{2}$
and the piecewise solid line has $\frac{D}{|J|} = 2.5$. The vertical dashed
line gives the value of the magnetic field where the first excited state
in the ferromagnetic phase A changes.   } 
	    \label{fig_2} 
	    
\end{center}
\end{figure}


\begin{figure} 
\begin{center}
\includegraphics[width=5cm,height= 6.5cm,angle= 0]{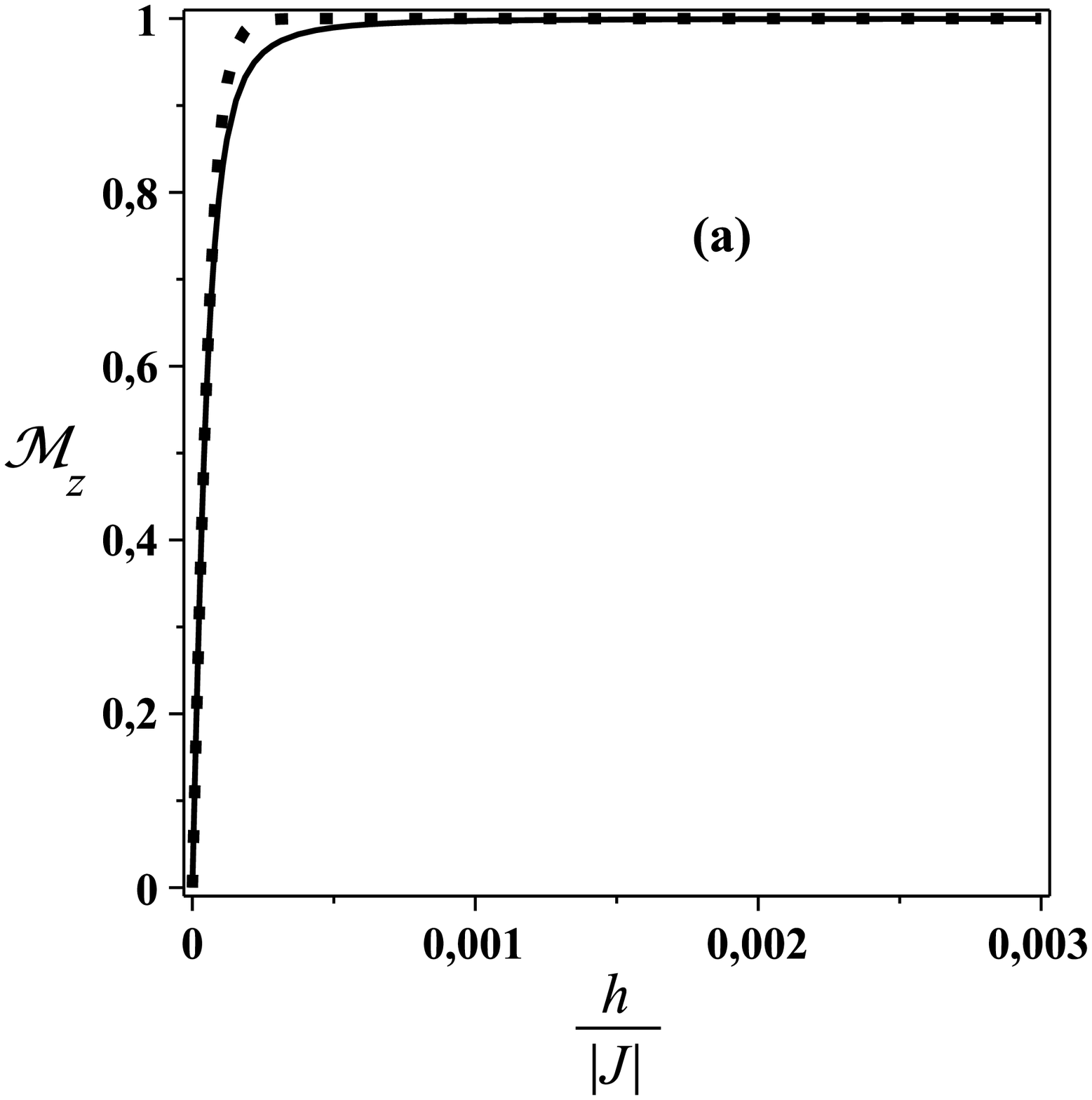} 
\includegraphics[width =4cm,height= 6.5cm,angle= 0]{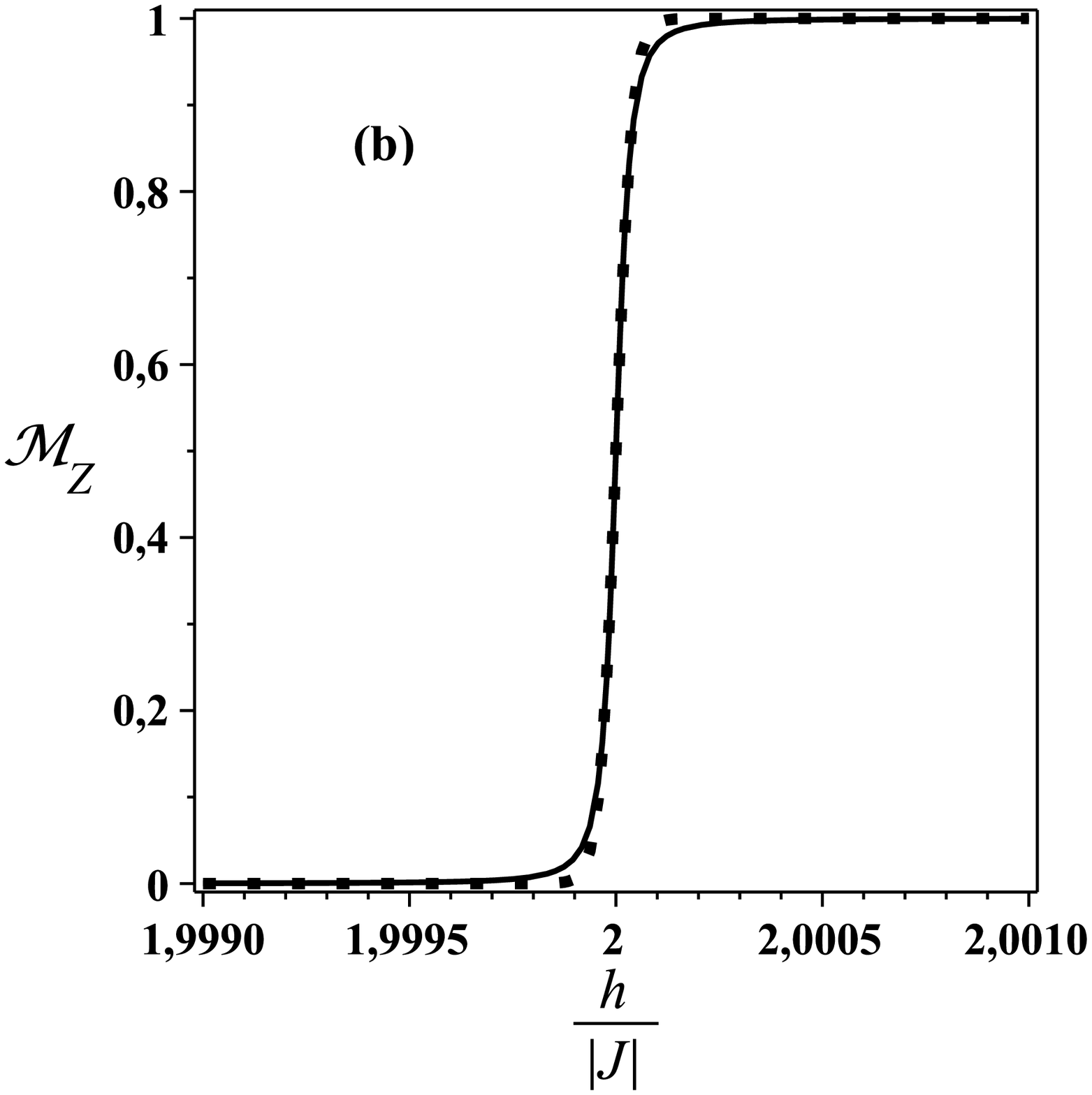} 
\includegraphics[width=6cm,height= 6.5cm,angle= 0]{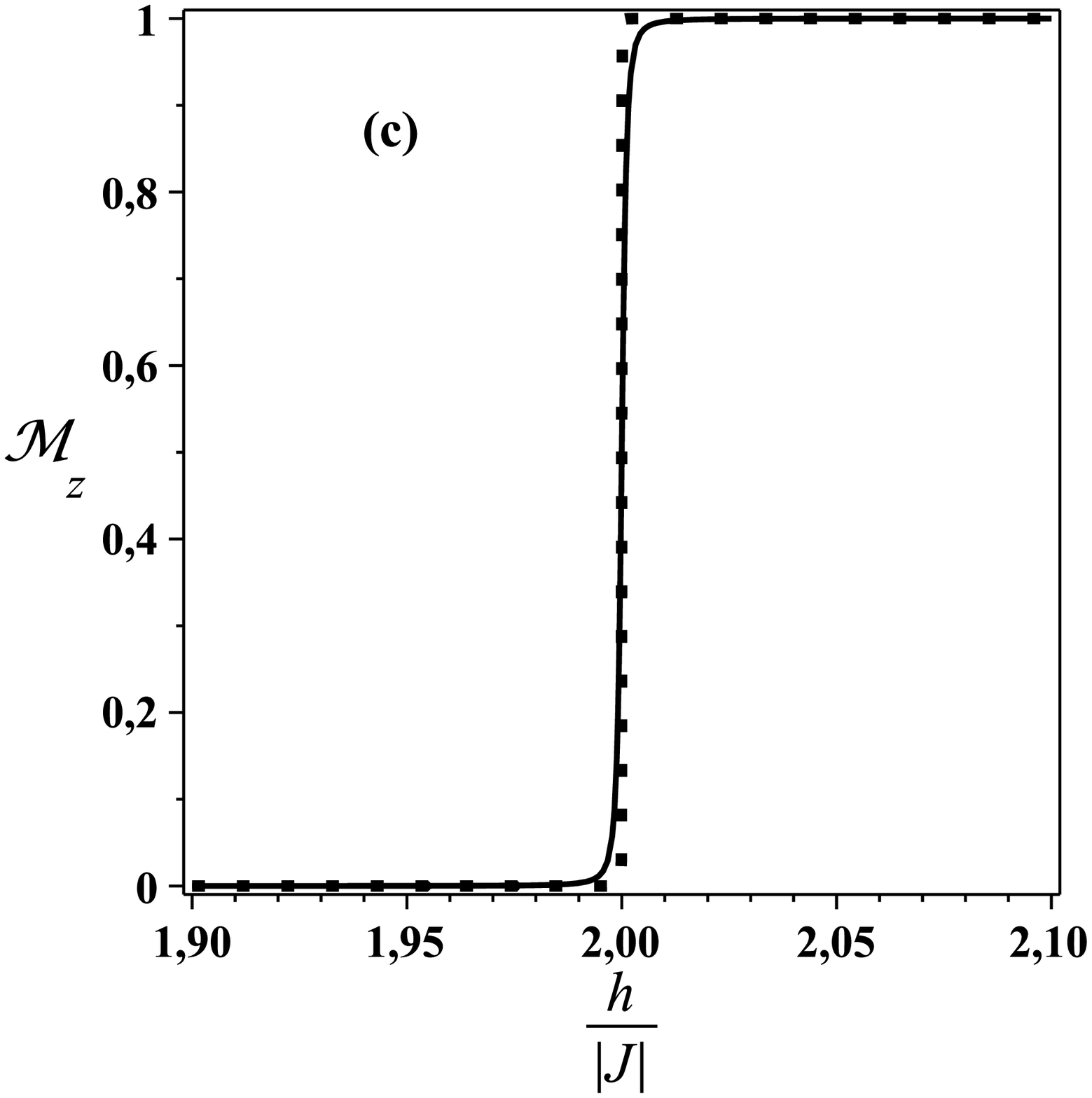} 
\end{center}
\vspace{-1cm}
	\caption{ The exact and approximate expressions of 
${\mathcal M}_z$ of the ferromagnetic  spin-1 model (\ref{1})/(\ref{5})
versus $\frac{h}{|J|}$. In $(a)$ we have $\frac{D}{|J|} = -2$
at $\frac{kT}{|J|} = 0.266$. The solid (dotted) line is its exact 
(approximate) expression.  In $(b)$ we have $\frac{D}{|J|} = 2.5$
at $\frac{kT}{|J|} = 0.0665$. The solid (dotted) line is its exact 
(approximate) expression. In $(c)$ we compare the exact expressions
of ${\mathcal M}_Z$  with $\frac{D}{|J|} = 2.5$  at temperatures
$\frac{ k T}{|J|} = 0.11$ (solid line) and $0.0665$ (dotted line).
}  \label{fig_3}  
 \end{figure}


\begin{figure} 
\begin{center}
\includegraphics[width= 6cm,height= 7.5cm,angle= 0]{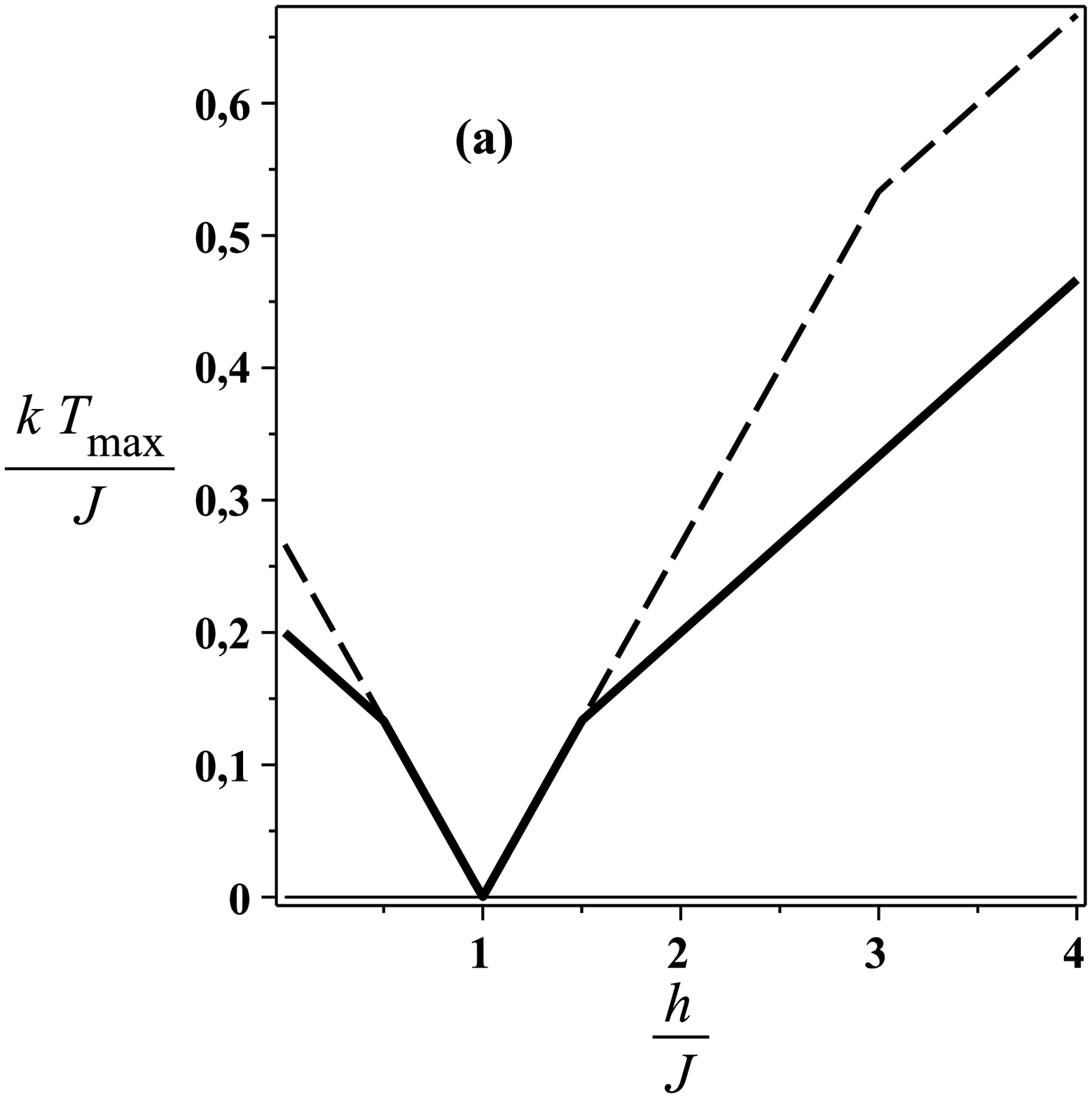} 
 \hspace{0.4cm}
\includegraphics[width =6cm,height= 7.5cm,angle= 0]{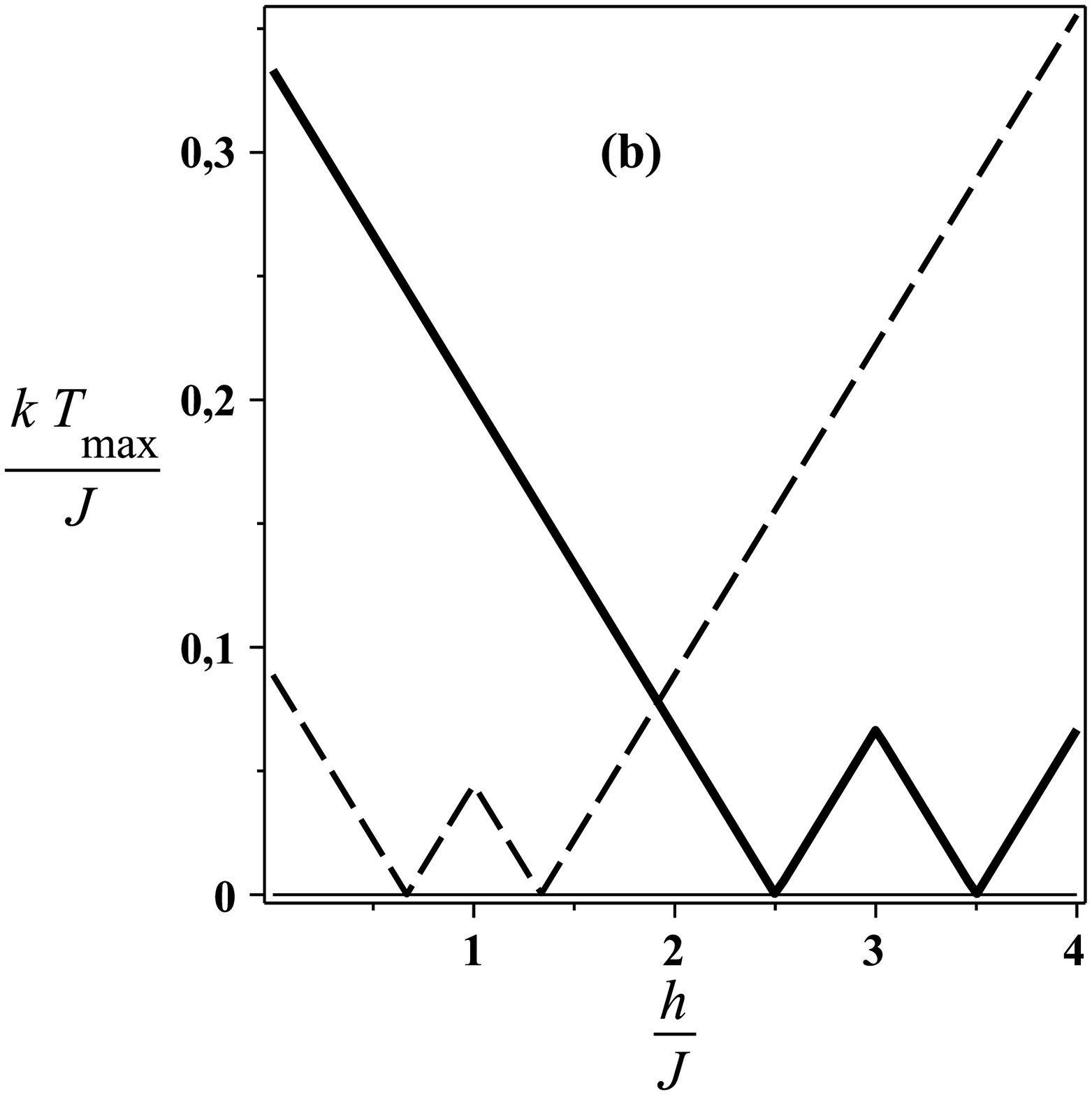}
\hspace{0.1cm}
\end{center}
\vspace{-0.7cm}
	\caption{ The functions 
$\frac{k T_{max}}{J} \times  \frac{h}{J}$ for the AF classical
spin-1 model. In $(a)$ we have the curves with $\frac{D}{J} = -2$
(dashed line) and $\frac{D}{J} = - \frac{1}{2}$ (solid line),
and in $(b)$  $\frac{D}{J} = \frac{1}{3}$ (dashed line) and
$\frac{D}{J} = 2.5$ (solid line)
}   \label{fig_4}   
\end{figure}


\begin{figure} 
\begin{center}
\includegraphics[width=6cm,height= 7.5cm,angle= 0]{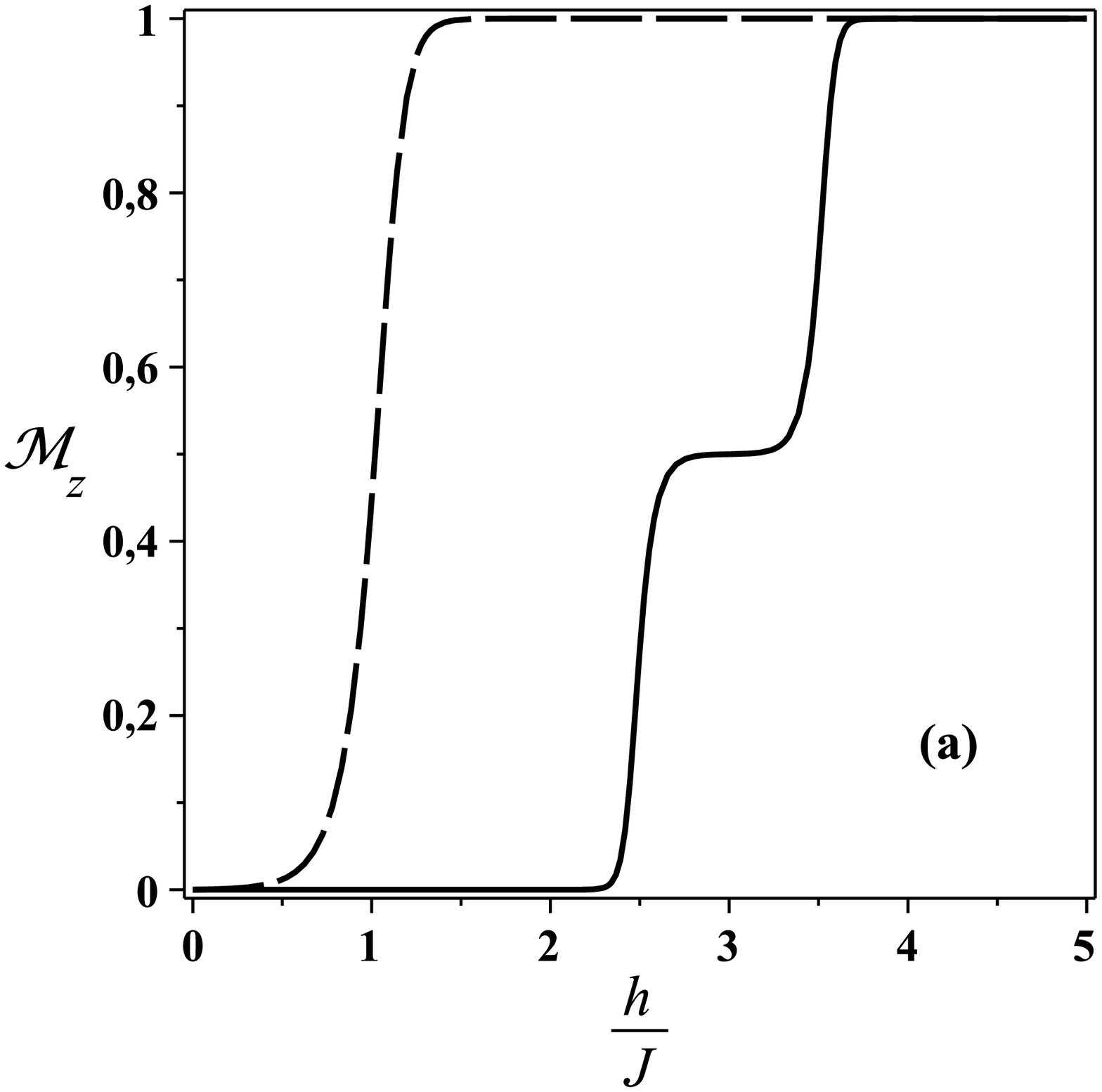} 
\includegraphics[width =6cm,height= 7.5cm,angle= 0]{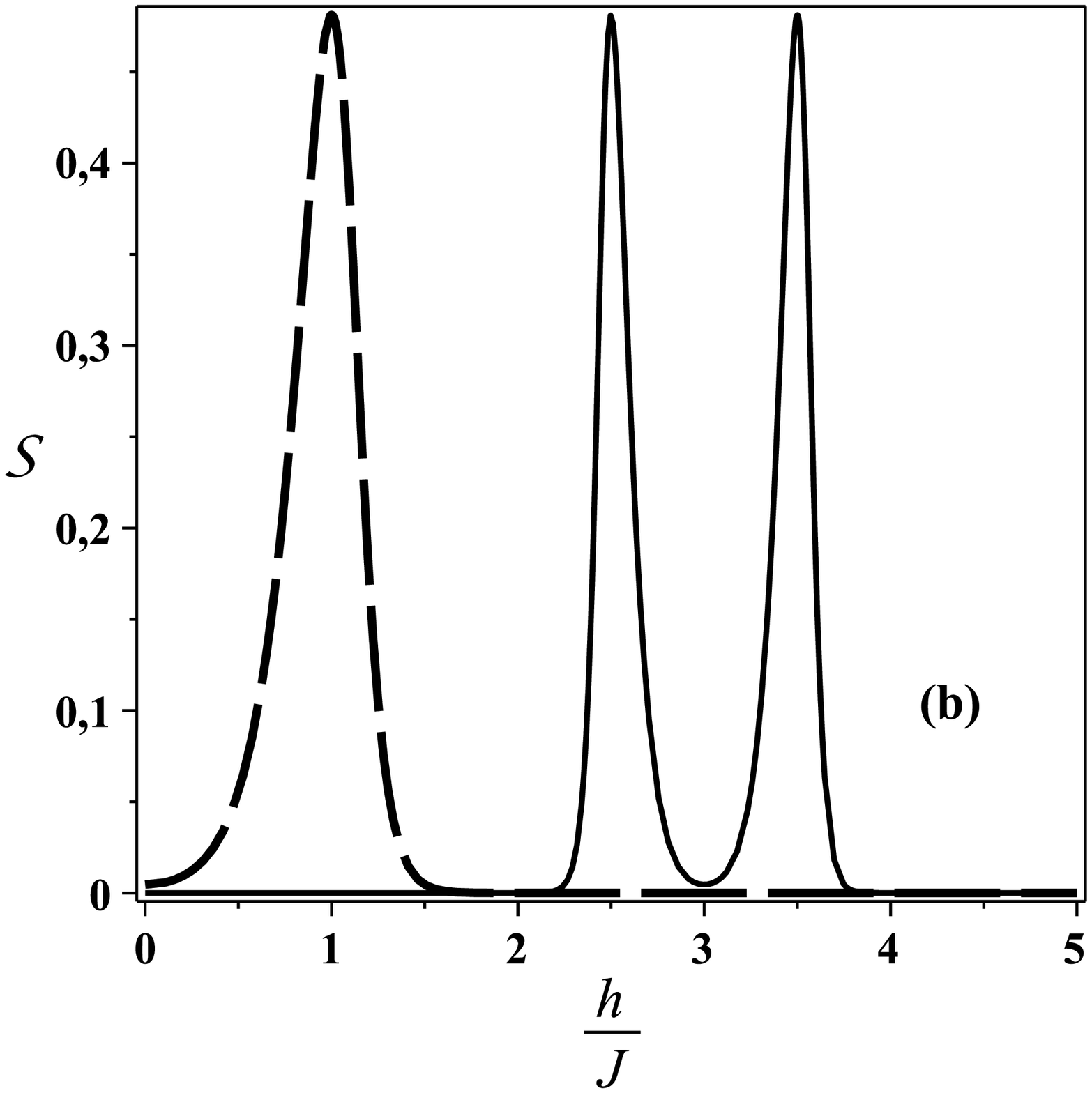} 
\end{center}
\vspace{-0.7cm}
	\caption{
The thermodynamic functions ${\mathcal M}_z$  and   
${\mathcal S}$  of the AF spin-1 model (\ref{1})/(\ref{5}) are 
plotted versus  $\frac{h}{J}$ with $\frac{D}{J} = -2$ and $2.5$. 
$(a)$ ${\mathcal M}_z$ as a function of  $\frac{h}{J}$ 
with $\frac{D}{J} = -2$ at $\frac{T}{|J|} = 0.266$ (long-dashed line) 
and with $\frac{D}{J} = 2.5$ at $\frac{T}{|J|} = 0.0665$ (solid line). 
The entropy per site, ${\mathcal S}$, is plotted  in ($b$)
for the same set of parameters: $\frac{D}{J} = -2$ at 
$\frac{T}{|J|} = 0.266$ (long-dashed line) and  
$\frac{D}{J} = 2.5$ at $\frac{T}{|J|} = 0.0665$ (solid line).
}   \label{fig_5}   
\end{figure}


\begin{figure} 
\begin{center}
\includegraphics[width=6cm,height= 7.5cm,angle= 0]{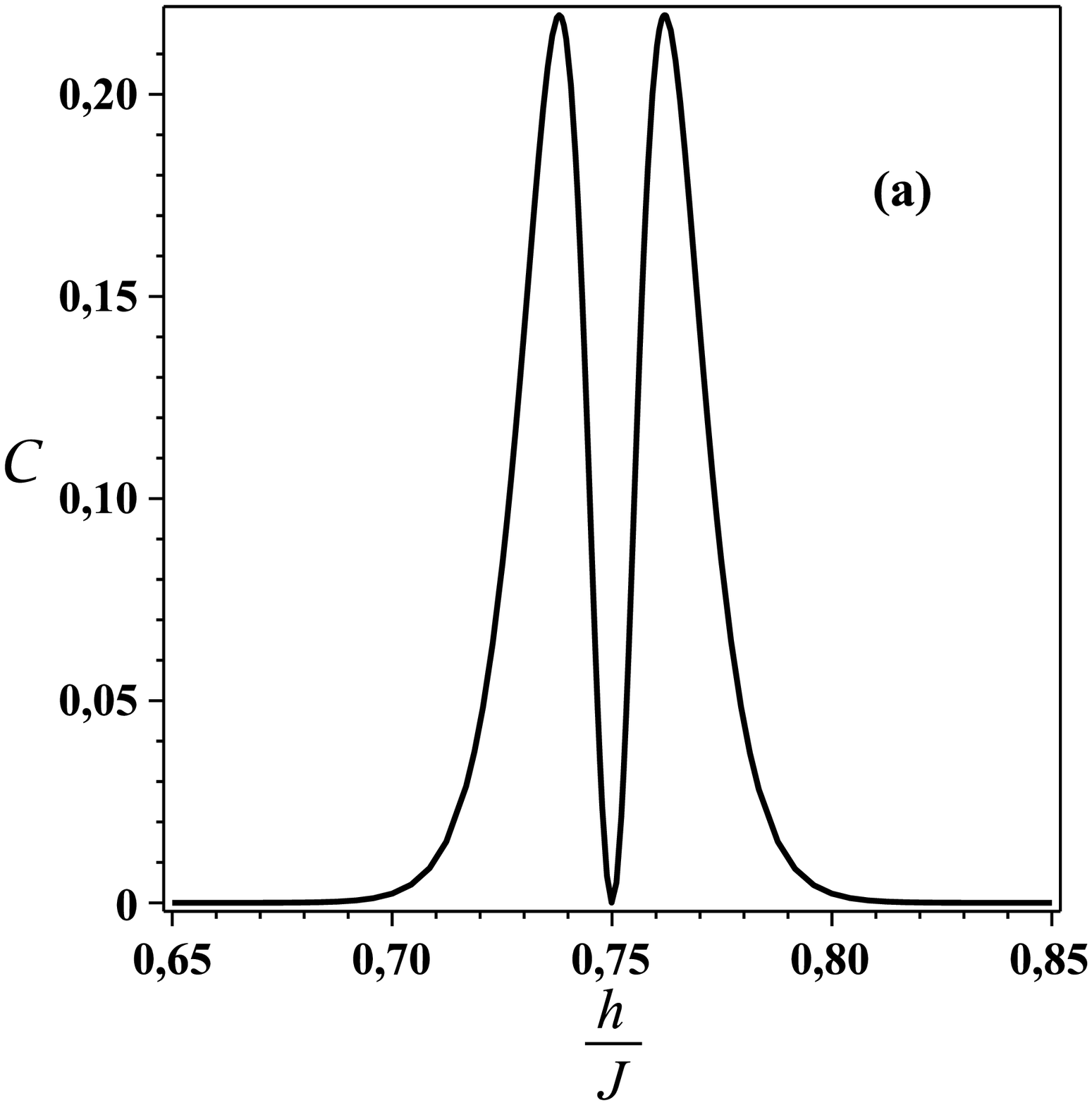} 
\hspace{1cm}
\includegraphics[width=6cm,height= 7.5cm,angle= 0]{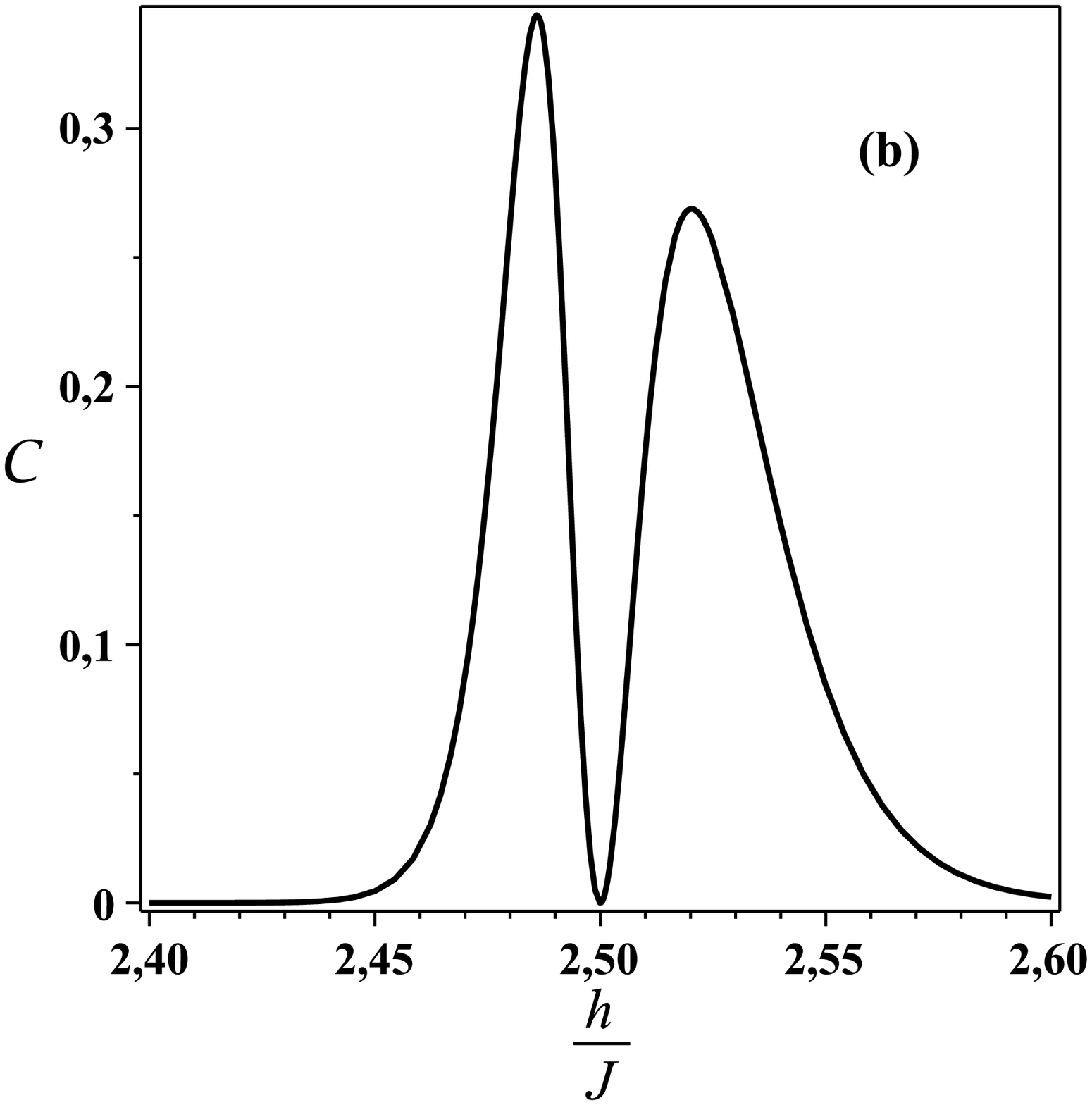}
\end{center}
\vspace{-0.7cm}
\caption{Comparison of the behavior of the function 
${\mathcal C} \times \frac{h}{J}$
under  the transitions $\mbox{G} \rightleftharpoons \mbox{E}$  
and $\mbox{A} \rightleftharpoons \mbox{E}$ at low temperature. 
In $(a)$ we  have the transition $\mbox{G} \rightleftharpoons \mbox{E}$ 
with $\frac{D}{J} = 0.25$. In $(b)$ is plotted the specific heat per site
of the transition $\mbox{A} \rightleftharpoons \mbox{E}$ with
$\frac{D}{J} = 2.5$. For both curves we have
 $\frac{k T}{J} = 0.01$.
}   \label{fig_6}   
\end{figure}


\begin{table} [ht] 
\begin{center}
\begin{tabular}{||c|c|c|c||}    
\hline
\hline
\rule{0em}{2.5ex} \rule{1cm}{0ex} & \rule{1cm}{0ex}  & \rule{1cm}{0ex} & \rule{1cm}{0ex} \\
$\frac{D}{|J|}$  & $\frac{k T}{|J|}$ &  A & $\frac{\Delta h}{|J|}$   \\
   &  &  &  \\
\hline
\hline
\rule{0em}{3ex} -2   & 0.06  & $2.9 \cdot 10^{14}$  & $1.3 \cdot 10^{-15}$ \\
\rule{0em}{3ex}    & 0.133  & $3.39  \cdot 10^{6}$  &  $1.8  \cdot 10^{-7}$ \\
\rule{0em}{3ex}    & 0.266  & $1.842  \cdot 10^3 $   & $6.7 \cdot  10^{-4}$  \\ 
\hline
  \rule{0em}{3ex} $- \frac{1}{2}$   & 0.0665  &$ 1.15 \cdot 10^{13}$ & $ 2.9 \cdot 10^{-14}$ \\
  \rule{0em}{3ex}    & 0.1995 & $2.25 \cdot 10^4 $ & $ 4 \cdot 10^{-6}$ \\ 
\hline	
  \rule{0em}{3ex} 2.5   & 0.04  & $4.9 \cdot 10^5$  & $10^{-6}$ \\
  \rule{0em}{3ex}    &  0.0665 & $3.42 \cdot 10^3$  & $10^{-3}$ \\ 
\hline
\hline
\end{tabular}
\end{center}
\caption{
The values of the parameter $a$ in eqs.(\ref{24}) and 
(\ref{27}) for several values of $\frac{D}{|J|}$ and different values 
of temperature. $\frac{\Delta h}{|J|}$ is the 
interval of variation of the magnetic field in which the
function ${\mathcal M}_z$ varies 
continuously between ${\mathcal M}_z = 0$
and ${\mathcal M}_z= 1$.
}     \label{tab_1}
\end{table}


\begin{thebibliography}{}

\bibitem{kramers1} H.A. Kramers and G.H. Wannier, Phys. Rev. 
{\bf 60} (1941) 252.

\bibitem{kramers2} H.A. Kramers and G.H. Wannier, Phys. Rev. 
{\bf 60} (1941) 263.

\bibitem{baxter} R.J. Baxter, "{\it Exactly Solved Models
in Statistical Mechanics}", Academic Press (1989), section 2.1.

\bibitem{simon} J. Simon {\it et al}., Nature {\bf 472}, 
04/21/2011, p. 307, and references therein.

\bibitem{blume} M. Blume, Phys. Rev. {\bf 141} (1966) 517.

\bibitem{capel} H.W. Capel, Physica (Utrecht) {\bf 32} (1966) 966;
{\bf 33} (1967) 295.

\bibitem{aydiner} E. Aydiner and C. Aky\"uz, Chin. Phys. Lett.
{\bf 22} (2005) 2382.

\bibitem{chen} X.Y. Chen {\it et al},  J. of Mag. and Mag. Mat 
{\bf 262} (2003) 258.

\bibitem{mancini}  F. Mancini, Europhys. Lett. {\bf 70} (2005) 485.

\bibitem{mancini2008} F. Mancini and F.P. Mancini, Cond. Matt. Phys. 
{\bf 11} (2008) 543.

\bibitem{winder} W.A. Moura-Melo {\it et al}., Phys. A {\bf 322}
(2003) 393.  Please notice a misprint in eq.(28) of this reference: the correct
coefficient of the term $(p^+ + p^- + r^-)$, inside 
the square root, is $\frac{8}{3}$.

\bibitem{krinsky} S. Krinsky and D. Furman, Phys. Rev. {\bf B11}
(1975) 2602. 

\bibitem{litaiff} F. Litaiff, R. de Sousa and N.S. Branco, Solid 
State Communication {\bf 147} (2008) 494.

\bibitem{ekiz}  C.Ekiz, H. Yaraneri, J. of Mag. and Mag. Mat 
{\bf 318} (2007) 49, and references therein.

\bibitem{narumiPhysB}  Y. Narumi {\it et al}., Phys. {\bf B246} 
(1998) 509.

\bibitem{narumiJMMM} Y. Narume {\it et al}., J. of Mag. and Mag. Mat 
{\bf 177-181} (1998) 685.

\bibitem{goto} T. Goto  {\it et al.}. Phys. {\bf B294-295} (2001) 43.


\bibitem{luz} D. Luz, R.R. dos Santos, Phys. Rev. {\bf B54} (1996) 1302.

\bibitem{moutinho} M. Moutinho, E.V. Corrêa Silva and M.T. Thomaz, 
Phys. {\bf A336} (2004) 477.


\bibitem{chain_m} O. Rojas, S. M. de Souza, and M. T. Thomaz,
 J. Math. Phys. {\bf 43} (2002) 1390.
 
\bibitem{isingSz2} M.T. Thomaz and O. Rojas, Condensed Matter 
Physics {\bf 15} (2012) 13706: 110 \\
\mbox{[dx.doi.org/10.5488/CMP.15.13706, http://www.icmp.lviv.ua/journal]}.


\bibitem{physA2011} E.V. Corr\^ea Silva, S.M de Souza and M.T. 
Thomaz, Phys. {\bf A390} (2011) 3108.

\bibitem{schaum} M.R. Spiegel, "{\it Mathematical Handbook of
Formula and Tables}", Schaum's Outline Series, Singapore (1990),
page 32.

\bibitem{meio} The factor $\left(\frac{1}{2}\right)$ in the relation between
${\mathcal M}_z$ and ${\mathcal W}_1$ comes from  the symmetric
form of Hamiltonian (\ref{5}) in the sites $i$ and $i+1$.

\bibitem{alcaraz} F.C. Alcaraz, S.R. Salinas and W.F. Wreszinski, 
Phys. Rev. Lett. {\bf 75} (1995) 930.

\bibitem{OYA}  M. Oshikawa, M. Yamanaka and I. Affleck, Phys. Rev.
Lett. {\bf 78} (1997) 1984.

\end{thebibliography}
\end{document}